\documentclass[twocolumn,showpacs,prc,aps,floatfix,reprint]{revtex4-1}

\usepackage{graphicx}
\usepackage{dcolumn}
\usepackage{bm}
\usepackage{hyperref}
\usepackage{amssymb}
\usepackage{braket}
\usepackage{graphicx}
\usepackage{dcolumn}
\usepackage{amsmath}
\usepackage{amsfonts}
\usepackage{color}
\usepackage{bm}
\usepackage[capitalise]{cleveref}
\usepackage{cleveref}
\usepackage{siunitx}

\begin{document}

\preprint{APS/123-QED}

\title{Single-particle structure of the semi-magic nucleus ${}^{90}$Zr  from a nonlocal \\dispersive optical model}

\author{R. A. Ramon$^{1}$, M. C. Atkinson$^{2,3}$, and W. H. Dickhoff$^1$}

\affiliation{${}^1$Department of Physics,
Washington University, St. Louis, Missouri 63130}
\affiliation{${}^2$Department of Physics,
Reed College, Portland, OR 97202}
\affiliation{${}^3$Department of Physics and Astronomy,
Haverford College, Haverford, PA 19041}

\date{\today}

\begin{abstract}
A nonlocal dispersive-optical-model (DOM) analysis has been carried out for neutrons and protons in the semi-magic nucleus $^{90}$Zr. Elastic-scattering angular distributions, total and reaction cross sections, single-particle energies, the neutron and proton numbers, the charge distribution, and the binding energy have been fitted to extract the neutron and proton self-energies both above and below the Fermi energy. 
The resulting spectroscopic factors and other DOM ingredients yield a good description of the $(e,e'p)$ cross sections when the open-shell proton system is described with an extension of the DOM that treats pairing.
The distinct difference between the open-shell proton system and a closed one is illustrated by the smooth transition from mostly full to mostly empty orbits. 

\end{abstract}

\maketitle

\section{Introduction}
\label{sec:intro}

The shell model, in which the nucleons fill certain orbitals, is well-suited to describe the structure of atomic nuclei. The best place to test this description is in or around (double) closed-shell nuclei. In the simplest picture, where residual interactions are neglected, the independent particle model (IPM) contains fully filled orbitals up to the Fermi level according to the Pauli principle, and those above it are empty. However, due to residual interactions ultimately linked to the underlying nucleon-nucleon interaction, there is a depletion of orbitals below the Fermi energy and filling of those above it. The precise amount of this depletion/filling continues to be a topic under investigation. The best tool to study this experimentally is the electron-induced proton knockout, $(e,e'p)$, reaction~\cite{Denherder:1988,Kramer:1989,Peter90,Lex90,Ingo91,Lapikas93,Pandharipande97}. At sufficiently high electron energy and momentum transfer, a bound proton within the struck nucleus can be ejected with enough energy such that a description within the distorted-wave impulse approximation (DWIA) can be expected to be applicable, so that depletion (and also filling) of orbits can be studied~\cite{Atkinson:2018}.

The picture emerging from experimental results of the $(e,e'p)$ reaction~\cite{Lapikas93} and theoretical work~\cite{Dickhoff04} suggests that double-closed-shell nuclei exhibit a reduction of the probability for valence proton removal of about 30\%.
This strength is redistributed above and below the Fermi energy by long- and short-range correlations. A successful description of these observations has been provided by the nonlocal version of the dispersive optical model (DOM)~\cite{Mahzoon:2014,Mahzoon:2017,atkinson2020dispersive} which yields an accurate description of elastic nucleon scattering data but also describes ground-state properties like the charge density. A critical validation of this approach was the accurate description of $(e,e'p)$ data for ${}^{40}$Ca~\cite{Atkinson:2018} and ${}^{48}$Ca~\cite{Atkinson:2019} without fitting to the corresponding experimental results.

While doubly closed-shell nuclei are the first systems to study, the nature of this depletion/filling is also widely investigated for rare isotopes~\cite{Aumann21}. Because rare isotopes predominantly represent open-shell nuclei, a study of the semi-magic nucleus ${}^{90}$Zr may yield useful information as there are ample experimental data and
a detailed study has been made of this nucleus with the $(e,e'p)$ reaction studied at Nikhef~\cite{Denherder:1988}. 
The proton filling of this nucleus is reminiscent of the smooth behavior of occupation numbers around the Fermi energy of a paired system~\cite{Allaart88}. We therefore consider the superfluid version of the Green's function method to include pairing correlations for the protons as a suitable extension of the dispersive optical model~\cite{Migdal:1967,Exposed!}.
 
Nuclei such as $^{90}$Zr are typically studied with emphasis on properties of the ground state and low-lying excitations like mean-field models~\cite{Bender:2003} and shell-model calculations~\cite{RevModPhys.92.015002}. 
However, these models are insufficient when considering the depletion/filling of orbitals below and above the Fermi energy. 
One natural avenue to study these effects is through microscopic and \textit{ab initio} methods. 
Although \textit{ab initio} methods are advancing at a rapid rate and can even describe properties of doubly-closed-shell nuclei as heavy as $^{208}$Pb~\cite{hu2022ab}, a consistent \textit{ab initio} description of the continuum is still limited to light nuclei. The importance of the scattering domain to clarify the presence of spectral strength associated with orbits fully occupied in the IPM has been documented in an earlier DOM analysis~\cite{Dussan:2014}. The link between the structure domain below the Fermi energy and the continuum above through a subtracted dispersion relation remains the hallmark of the DOM~\cite{Mahaux91,Dickhoff:2017,Dickhoff:2019,Exposed!}.
We note that recent traditional DOM calculations for ${}^{90}$Zr~\cite{Capote:2023} and a sequence of Zr isotopes~\cite{Capote:2026} are available employing local potentials.
These DOM applications mostly include scattering results and level structure and consider ${}^{90}$Zr a double closed-shell nucleus.
As shown in the present paper, treating $^{90}$Zr as a double closed-shell nucleus is an oversimplification as the quenching of the spectroscopic factor of the proton orbit on the Fermi surface is well documented experimentally~\cite{Denherder:1988} and can only be described by utilizing nonlocal DOM potentials and the extension to include pairing as presented here.

An additional motivation for studying ${}^{90}$Zr is the possibility of predicting its neutron skin defined as the difference between the point neutron and proton root-mean-squared (RMS) radii, $R_\mathrm{skin} = R_n - R_p$, of a finite nucleus. This nucleus has a particle number between ${}^{48}$Ca and ${}^{208}$Pb but a smaller asymmetry than ${}^{48}$Ca allowing further study of this important quantity that is well described by the DOM~\cite{calleya2025investigating} for ${}^{48}$Ca~\cite{adhikari2022precision} and ${}^{208}$Pb~\cite{adhikari2021accurate}. 

Thus, in this paper, we extend the DOM to treat open-shell semi-magic nuclei by including the appropriate pairing formalism. We analyze the single-particle structure of $^{90}$Zr after constraining the DOM self-energy to reproduce elastic-scattering data as well as energy levels, total binding energy, particle number, and charge density. We then concentrate on the 
$^{90}$Zr$(e,e'p)^{89}$Y data available from the Nikhef experiments~\cite{Denherder:1988}. 
We calculate $^{90}$Zr$(e,e'p)^{89}$Y momentum distributions using the distorted wave impulse approximation, as discussed in~\cite{Atkinson:2018}. We study the transitions to the levels nearest the Fermi energy, which can be considered discrete transitions after having applied the splitting of the strength across the Fermi energy implied by the pairing description. 

In Sec.~\ref{sec:theory} we review the underlying theory for the dispersive optical model (Sec.~\ref{sec:dom}) as well as the pairing formalism (Sec.~\ref{sec:pairing}). We present the result of the DOM fit in Sec.~\ref{sec:fit-results} and the $^{90}$Zr$(e,e'p)^{89}$Y reaction calculation is discussed in Sec.~\ref{sec:eep}. 
Conclusions and outlook are discussed in Sec.~\ref{sec:conclusion}.

\section{Theory}
\label{sec:theory}

Underlying the dispersive optical model, as well as our approach to pairing, is the many-body Green's function formalism~\cite{Exposed!,Migdal:1967}. Many of the expressions in this section are discussed in further detail in Ref.~\cite{Exposed!}.

\subsection{Dispersive optical model}
\label{sec:dom}

In the many-body Green's function formalism, the irreducible self-energy, $\Sigma^*(\bm{r},\bm{r}';E)$, is a complex one-body potential which, in principle, is comprised of an infinite set of Feynman diagrams describing the propagation of an interacting nucleon through a nucleus based on a Hamiltonian containing relevant two- and three-body interactions~\cite{Exposed!}. In the DOM formalism, this complex one-body potential is parametrized as an optical potential which naturally extends to negative (bound) energies within the Green's function framework~\cite{Mahaux91}. 
The analytic structure of the nucleon self-energy allows one to utilize a dispersion relation, which relates the real part of the self-energy at a given energy to a dispersion integral of its imaginary part over all energies. The energy-independent correlated Hartree-Fock (HF) contribution~\cite{Exposed!} is removed by employing a subtracted dispersion relation with the Fermi energy used as the subtraction point~\cite{Mahaux91}
  \begin{align}
    \mathrm{Re}\ \Sigma^*(\bm{r},\bm{r}';E) &= \mathrm{Re}\
    \Sigma^*(\bm{r},\bm{r}';\varepsilon_F) \label{eq:dispersion} \\ -
    \mathcal{P}\int_{\varepsilon_F}^{\infty} \!\! \frac{dE'}{\pi}&\mathrm{Im}\
    \Sigma^*(\bm{r},\bm{r}';E')\left[\frac{1}{E-E'}-\frac{1}{\varepsilon_F-E'}\right] \nonumber
    \\ + \mathcal{P} \! \int_{-\infty}^{\varepsilon_F} \!\!
    \frac{dE'}{\pi}&\mathrm{Im}\
    \Sigma^*(\bm{r},\bm{r}';E')\left[\frac{1}{E-E'}-\frac{1}{\varepsilon_F-E'}\right],
    \nonumber      
 \end{align}
 where $\bm{r}$ also implies relevant discrete quantum numbers and $\varepsilon_F$ is the average Fermi energy which separates the particle and hole domains, 
 \begin{align*}
 \varepsilon_F = \frac{1}{2}\left(E_0^{A+1}-E_0^{A-1}\right),
 \end{align*}
 and $E_0^{A\pm1}$ represent the ground-state energies of the $A\pm1$ nucleus~\cite{Exposed!}. 

Earlier work to include ground-state properties in the DOM demonstrated that it was not possible to reproduce the charge density without including a nonlocal imaginary part~\cite{Dickhoff:2010}.
We therefore implement a nonlocal representation of the self-energy following
 Ref.~\cite{Mahzoon:2014} where $\Sigma_{\text{HF}}(\bm{r},\bm{r'}) = \mathrm{Re}\
    \Sigma^*(\bm{r},\bm{r}';\varepsilon_F) $ and the imaginary part 
 $\mathrm{Im}\ \Sigma(\bm{r},\bm{r'};E)$ are parametrized, and Eq.~\eqref{eq:dispersion} generates the energy dependence of the
 real part. The HF term consists of a volume term, spin-orbit term,  and a wine bottle shape consistent with a microscopic analysis~\cite{Brida11}. The imaginary self-energy consists of volume, surface, and spin-orbit terms. 
 Nonlocality is represented using the Gaussian form as proposed in Ref.~\cite{Perey:1962}. 
 More details can be found in Ref.~\cite{atkinson2020dispersive}. The parameterization is presented in App.~\ref{app:params}.

To use the DOM self-energy for predictions, the parameters are fit through a weighted $\chi^2$ minimization of available elastic differential cross section data ($\frac{d\sigma}{d\Omega}$), analyzing power data ($A_\theta$),  reaction cross sections ($\sigma_r$), total cross sections ($\sigma_t$), charge density ($\rho_{\text{ch}}$), energy levels ($\varepsilon_{\ell j}$), particle number, the root-mean-square charge radius ($R_\mathrm{ch}$), and the energy of the ground-state. The scattering calculations are performed in a partial wave basis (specifying orbital and total angular momentum with $\ell$ and $j$, respectively) using $\Sigma_{\ell j}^*(r,r';E)$ as an optical potential in the framework of $R$-matrix theory~\cite{Baye:2010}.
 All calculations are done in a Lagrange basis with 30 mesh points, where Legendre polynomials with a matching radius of 12 fm are used for scattering calculations and Laguerre polynomials are used for bound-state calculations~\cite{Baye_review,Baye:2010}.

We employ the Dyson equation to obtain the Green's function, $G_{\ell j}(r,r';E)$, from the DOM self-energy, 
  \begin{align}
    G_{\ell j}(r,r';E) &= G_{\ell}^{(0)}(r,r';E) \nonumber \\ +
    \int dr_1 dr_2 r_1^2 r_2^2 &G_{\ell}^{(0)}(r,r_1;E)\Sigma_{\ell
    j}^*(r_1,r_2;E)G_{\ell j}(r_2,r';E) ,
    \label{eq:dyson}
 \end{align}
 where $G^{(0)}_{\ell}(r,r';E)$ corresponds to the free propagator (the Green's function when $\Sigma_{\ell j}^*(r_1,r_2;E)=0$)
 ~\cite{Exposed!}. The particle number, binding energy, and charge density are all obtained from the hole spectral density which corresponds to the imaginary part of the Green's functions, 
 \begin{equation}
    S^{(P,N)}_{\ell j}(r,r';E) = \frac{1}{\pi}\mathrm{Im}\ G^{(P,N)}_{\ell j}(r,r';E) ,
    \label{eq:spec}
 \end{equation}
with notation $P$ and $N$ for proton and neutron, respectively.
The single-particle density distribution can be calculated from the hole spectral function in the following way, 
 \begin{equation}
    \label{eq:charge}
    \rho^{(P,N)}(r) = \frac{1}{4\pi} \sum_{\ell j} (2j+1) \int_{-\infty}^{\varepsilon_F}dE\ S^{(P,N)}_{\ell j}(r;E),
 \end{equation}
 where $S^{(P,N)}_{\ell j}(r;E) = S^{(P,N)}_{\ell j}(r,r;E)$ is the diagonal of the hole spectral density.

 The spectral strength for a given $\ell j$ partial wave can be found by summing (integrating) the spectral function according to
 \begin{equation}
    S_{\ell j}^{(P,N)}(E) = \int_0^\infty S_{\ell j}^{(P,N)}(r;E)r^2dr,
    \label{eq:strength}
 \end{equation}
 when $E<\varepsilon_F$.
 The spectral strength $S_{\ell j}^{(P,N)}(E)$ is the contribution at energy $E$ to the occupation from all orbitals with orbital angular momentum $\ell$ and total angular momentum $j$.  
 \begin{figure}[t]
            \includegraphics[width=\columnwidth]{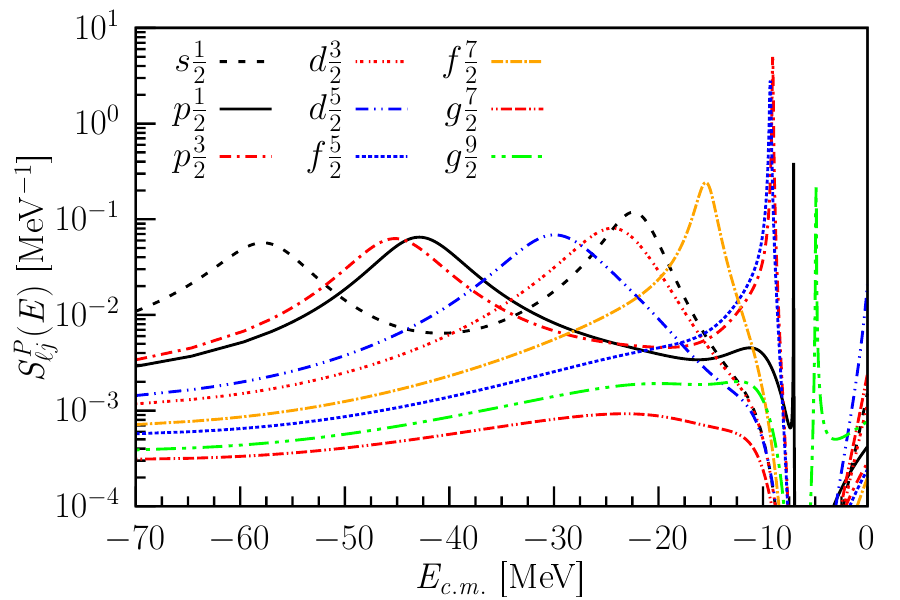}
      \caption{Proton spectral functions for a representative set of $lj$ shells in $^{90}$Zr without the treatment of pairing. The sharp peaks above and below the Fermi energy (-6.754 MeV) are for illustration purposes only as these peaks are treated by solving Eq.~\eqref{eq:schrodingerq} with the corresponding normalization of Eq.~\eqref{eq:sf}.}
   \label{fig:spectral_p}
\end{figure}
 Figure~\ref{fig:spectral_p} shows the spectral strength for a representative set of proton shells in $^{90}$Zr that would be considered fully occupied in the IPM.
 In addition, the $g\frac{9}{2}$ strength with main peak above the Fermi energy is displayed. These results do not yet include the fragmentation due to pairing. The peaks in
 Fig.~\ref{fig:spectral_p} correspond to the binding energy of the appropriate IPM orbital. 
 While $S_{\ell j}(E)$ is concentrated at these energy levels, there is still substantial strength fragmented over a wide range of energies. 

The wave functions of these quasihole/particle states can be obtained by transforming the Dyson equation into a nonlocal Schr\"{o}dinger-like equation by disregarding the imaginary part of $\Sigma_{\ell j}^*(r,r';E)$,
 \begin{align}
T_\ell \psi_{\ell j}(r) + \!\!
    \int_0^\infty \!\! \Sigma^*_{\ell j}(r,r';\varepsilon_{\ell j}^n)\psi_{\ell j}^n(r')r'^2dr' = \varepsilon_{\ell j}^n\psi_{\ell j}^n(r),
    \label{eq:schrodingerq}
 \end{align}
 where $T_\ell$ is the radial component of the kinetic energy operator in spherical coordinates, including the centrifugal term.
 The wave function, $\psi_{\ell j}^n(r)$, is the overlap between the $A$ and $A-1$ systems and the corresponding energy, $\varepsilon_{\ell j}^n$, is the energy required to remove the bound eigenstate with the particular quantum numbers $\ell j$.
 The value of $n$ distinguishes different solution, starting with $n = 0$ for the lowest one, 
 \begin{equation}
    \psi^n_{\ell j}(r) = \bra{\Psi_n^{A-1}}a_{r \ell j}\ket{\Psi_0^A}, \qquad \varepsilon_{\ell j}^n = E_0^A - E_{n\ell j}^{A-1} .
    \label{eq:wavefunction}
 \end{equation}

 When solutions to Eq.~\eqref{eq:schrodingerq} are found near the Fermi energy where there is no imaginary part of the self-energy (although a small imaginary component is utilized in practice), the normalization of the quasihole state generates the corresponding spectroscopic factor,
 \begin{equation}
    \mathcal{Z}^n_{\ell j} = \bigg(1 - \frac{\partial\Sigma_{\ell j}^*(\alpha_{qh},\alpha_{qh};E)}{\partial E}\bigg|_{\varepsilon_{\ell j}^n}\bigg)^{-1},
    \label{eq:sf}
 \end{equation}
 where $\alpha_{qh}$ corresponds to the quasihole state that solves Eq.~\eqref{eq:schrodingerq}. 
 This type of normalization loses its meaning as soon as the imaginary part becomes significant as the strength becomes correspondingly broadened as shown in Fig.~\ref{fig:spectral_p}.
 The quasihole peaks in Fig.~\ref{fig:spectral_p} thus become narrower as they approach $\varepsilon_F$, which is a consequence of the imaginary part of the irreducible self-energy vanishing when reaching this energy.
 In fact, the last mostly occupied proton level, $1p\frac{1}{2}$ (at $-7.07$ MeV) in Fig.~\ref{fig:spectral_p} exhibits spectral strength that is essentially a delta function peaked at its energy level according to Eq.~\eqref{eq:schrodingerq}.
 For these orbitals, the strength of the spectral function at the peak corresponds to the spectroscopic factor in Eq.~\eqref{eq:sf}. This spectroscopic factor is the very same we employ in the $(e,e'p)$ calculations discussed in Sec.~\ref{sec:eep} (see also Refs.~\cite{Atkinson:2018,Atkinson:2019}).

\subsection{Pairing}
\label{sec:pairing}

For a semi-magic nucleus with an even number of protons, ${}^{90}$Zr in this case, the single-particle strength of levels near the Fermi energy is split above and below the Fermi energy. This is evident from the experimental states reached with the stripping reaction, $^{90}$Zr($^{3}$He,$d)^{91}$Nb~\cite{knopfle1970reaction}, for information above the Fermi energy, and the electron-induced proton knockout reactions, $^{90}$Zr($e,e'p$)$^{89}$Y~\cite{Denherder:1988} for information below the Fermi energy. The $(e,e'p)$ reactions are considered the cleanest probe to study the single-particle properties, as the electron interacts weakly with the nucleus and the knocked out proton has sufficient energy of about 100 MeV to allow a DWIA treatment of the reaction~\cite{Atkinson:2018}.
Hadronic probes depend on their in-medium interaction that is much less well known thereby introducing additional uncertainties~\cite{Aumann21}. 

The proton shells in ${}^{90}$Zr most relevant for low-energy excitations are the $1p\frac{1}{2}$ and $0g\frac{9}{2}$ orbits.
These levels are close together and the $1p\frac{3}{2}$ and $0f\frac{5}{2}$ levels are also nearby. 
The attraction between pairs of protons coupled to angular momentum $J = 0$ rearranges the simple shell filling such that a smooth occupation of these orbits yields a preferred ground state. 
Such pairing correlations can be described by the Green's function method when a self-consistent formulation is adopted including the concept of anomalous propagators~\cite{Exposed!,Migdal:1967}.
When other correlations are important, there is a two-step procedure that can incorporate these effects. 
In infinite nuclear or neutron matter, this includes taking care of short-range correlations first, generating a so-called normal propagator which is then further renormalized by the solution of the gap equation~\cite{Ding:2016} which even at the BCS level splits the single-particle strength near the Fermi energy into partial occupation and emptiness~\cite{Exposed!}.

Such a treatment can also be implemented in the DOM analysis of ${}^{90}$Zr.
The usual DOM approach is to treat the $1p\frac{1}{2}$ orbits completely filled and the $0g\frac{9}{2}$ one completely empty as generated by the $\Sigma_{\textrm{HF}}$ contribution in Eq.\eqref{eq:dispersion}.
Scattering cross sections can then be adequately described when the imaginary part and its corresponding dynamic real part are included.
Even the charge density can be reasonably described as we show in Sec.~\ref{sec:fit-results}.

Nevertheless, the pairing correlations are sufficiently strong so that the splitting of the single-particle strength of the aforementioned orbits is essential.
This is explicitly probed by the $(e,e'p)$ reaction reported in Ref.~\cite{Denherder:1988} so that our approach can be tested by directly comparing to these data.
The step to include pairing effects has been presented in Ref.~\cite{Migdal:1967} and will be adopted here using the same notation.
This treatment distinguishes between the particle and hole part of the propagator and accordingly solves the coupling between the normal and anomalous propagators separately.
This separation employs the chemical potentials for adding and removing pairs of protons, $\mu_2^+$ and $\mu_2^-$, respectively, with
\begin{equation}
\mu^{+}_{2} = \frac{1}{2}\left[E_0(N,Z+2) - E_0(N,Z)\right]
\label{eq:muplus}
\end{equation}
\begin{equation}
\mu^{-}_{2} = \frac{1}{2}\left[E_0(N,Z) - E_0(N,Z-2)\right],
\label{eq:muminus}
\end{equation}
where $E_0(N, Z)$ is the ground-state binding energy of a nucleus with $N+Z$ nucleons. 
The quasiparticle energies then read
\begin{equation}
    E^{+}_{1\lambda} = \mu^{+}_2 +\sqrt{(\Delta^{+}_{\lambda})^{2} +(\varepsilon_\lambda - \mu^{+}_{2})^{2}} \equiv \mu^{+}_2 + E^{+}_{\lambda}
    \label{eq : E_plus1_lambda}
\end{equation}
and
\begin{equation}
    E^{-}_{1\lambda} = \mu^{-}_2 -\sqrt{(\Delta^{-}_{\lambda})^{2} +(\varepsilon_\lambda - \mu^{-}_{2})^{2}} \equiv \mu^{-}_2 - E^{-}_{\lambda}
    \label{eq : E_minus1_lambda},
\end{equation}
where $\varepsilon_\lambda$ refers to the single-particle energy from the initial DOM step, possibly different gap parameters $\Delta^\pm_\lambda$ are considered, and we have identified the signs in front of $E^\pm_\lambda$ as the ones relevant for an even proton number~\cite{Migdal:1967}.

For an even $Z$ nucleus, a simple estimate of the proton pairing energy $\Delta$ is obtained from~\cite{Bohr-Mottelson} as
 \begin{align}
 \begin{split}
    \Delta = \frac{1}{4}[E_{0}(N,Z-2) - 3E_{0}(N,Z-1)\\ + 3E_{0}(N,Z) - E_{0}(N,Z+1)] .
\end{split}
    \label{eq:E_pairing}
\end{align}
We use this ansatz with the same gap applying to both Eqs.~\eqref{eq : E_plus1_lambda} and ~\eqref{eq : E_minus1_lambda} to split the energies of the 1$p\frac{1}{2}$, 1$p\frac{3}{2}$, 0$f\frac{5}{2}$ and 0$g\frac{9}{2}$ levels calculated in the first DOM step from Eq.~\eqref{eq:schrodingerq}. Thus, with the inclusion of this pairing extension of the DOM we can now accurately describe the splitting of open-shell energy levels of semi-magic nuclei. 

Experimental data confirm fragments of 1$p\frac{1}{2}$ and 0$g\frac{9}{2}$ both above and below the Fermi energy~\cite{knopfle1970reaction,den1988single}. 
Spectroscopic factors corresponding to these single-particle levels must correspondingly be separated into two parts~\cite{Migdal:1967}. 
The expressions in~\cite{Migdal:1967} are correct when $\Delta^{+}_{\lambda}$ = $\Delta^{-}_{\lambda}$ and $E^{+}_{\lambda}$ = $E^{-}_{\lambda}$. 
In the present analysis, $E^{+}_{\lambda}$ $\neq$ $E^{-}_{\lambda}$ and $\Delta^{+}_{\lambda}$ = $\Delta^{-}_{\lambda} = \Delta$ . 
As a result, these spectroscopic factors are split according to

\begin{align}
\mathcal{Z}_{\text{lower}} &= \mathcal{Z} \times \frac{f_{\mathrm{L}}}{f_{\mathrm{L}} + f_{\mathrm{U}}}
\label{eq:zlower}
\\[6pt]
\mathcal{Z}_{\text{upper}} &= \mathcal{Z} \times \frac{f_{\mathrm{U}}}{f_{\mathrm{L}} + f_{\mathrm{U}}}
\label{eq:zupper},
\end{align}
with $\mathcal{Z}$ the DOM spectroscopic factor from the first step.
The lower, $f_L$ and upper, $f_U$ weighting factors are given by
\begin{align}
f_{\mathrm{L}}
&=
\frac{1}{2}
\left(
1 - \frac{\varepsilon - \mu^{-}_{2}}{\sqrt{
\left(E_{\lambda}^{-}\right)^2 + \Delta^2
}}
\right)
\\[10pt]
f_{\mathrm{U}}
&=
\frac{1}{2}
\left(
1 + \frac{ \varepsilon - \mu^{+}_{2}}{\sqrt{
\left(E_{\lambda}^{+}\right)^2 + \Delta^2
}}
\right).
\end{align}
The present implementation of pairing, therefore, only uses a single parameter, $\Delta$, for all four proton orbits and the experimental data given by Eqs.~(\eqref{eq:muplus}) and (\eqref{eq:muminus}).

\section{Results}
\label{sec:fit-results}

The functional form of the ${}^{90}$Zr self-energy is equivalent to that of ${}^{48}$Ca and ${}^{208}$Pb used in Refs.~\cite{Atkinson:2019} and~\cite{atkinson2020dispersive}, respectively. 
The parameters of the DOM self-energy (see App.~\ref{app:params}) are constrained through a weighted $\chi^2$ minimization. 
To minimize the $\chi^2$, the parameters are varied according to an algorithm based on the Powell method~\cite{Numerical}. In the remainder of this section, we present the observables produced by the resulting parameters.
    
\begin{figure}[t]
      \includegraphics[width=\columnwidth]{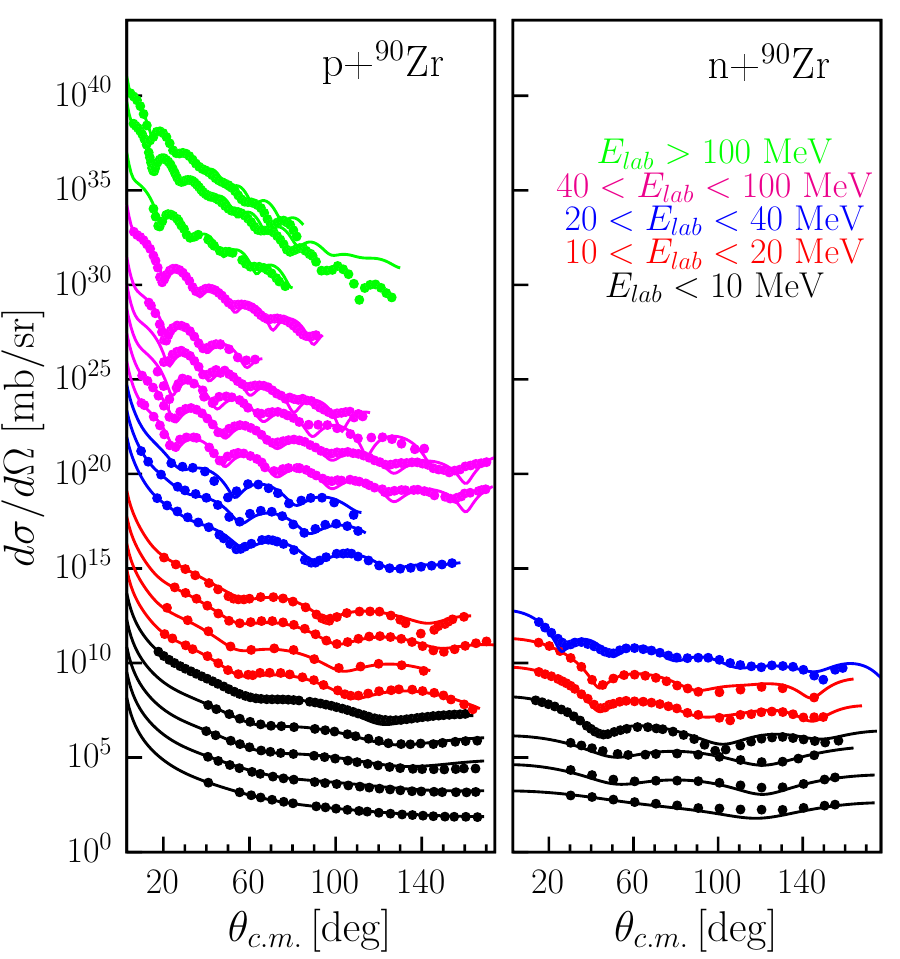}
      \caption{Angular distributions of the differential cross section of proton and neutron elastic scattering off $^{90}$Zr for the incident energy, $\mathrm{E_{lab}}$, ranging from  $5.574$ to $160.0$ MeV for protons and $1.50$ to $24.0$ MeV for neutrons. The data at each energy is offset by factors of 10 to visualize all the data (filled circles) and results (solid lines) simultaneously.}
   \label{fig:xsec_p_n}
\end{figure}

\begin{figure}[b]
    \includegraphics[width=\columnwidth]{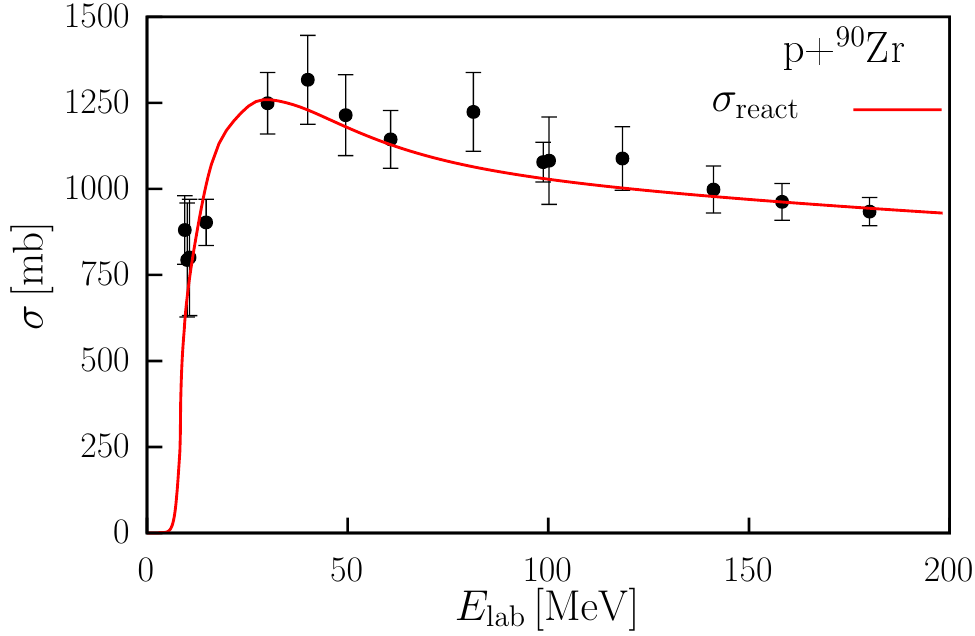}
      \caption{Proton reaction cross section for $^{90}$Zr. Experimental data points are from Ref.~\cite{auce2005reaction}}
   \label{fig:react_p}
\end{figure}

\begin{figure}[ht]
    \includegraphics[width=\columnwidth]{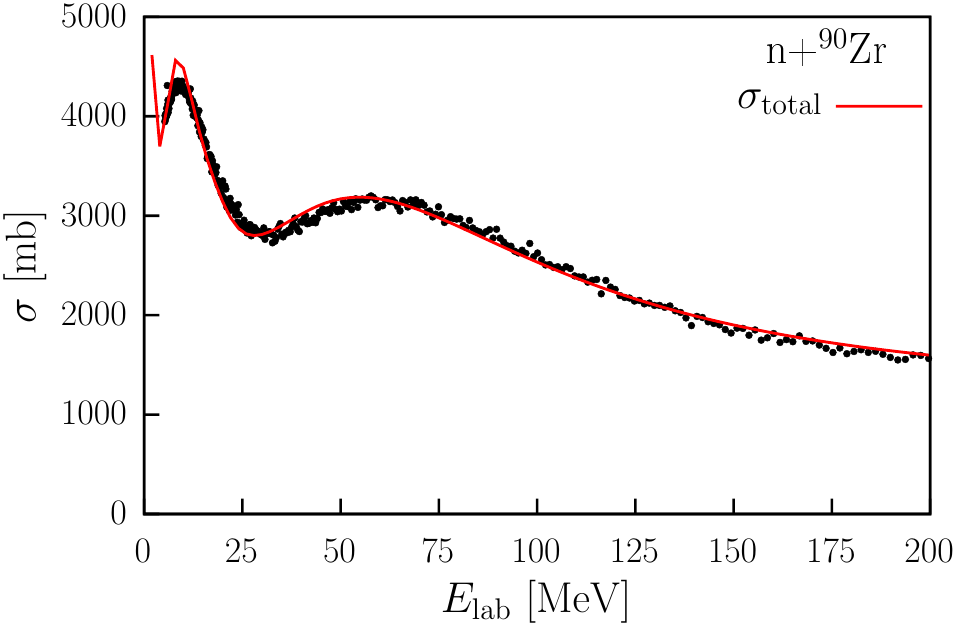}
      \caption{Total cross section for neutrons scattered off $^{90}$Zr. Experimental data points are from Ref.~\cite{finlay1993neutron}.}
   \label{fig:total_n}
\end{figure}

The angular distributions of the nucleon-nucleus elastic-scattering differential cross section for the energy range $5.574$ to $160.0$ MeV for protons and $1.50$ to $24.0$ MeV for neutrons are presented in Fig.~\ref{fig:xsec_p_n}. The energy dependence of the self-energy is responsible for the good quantitative agreement with the training data over this large energy range. 
The proton reaction cross section, which is an inclusive measurement of all inelastic processes that can result from the $p+^{90}$Zr collision, is displayed in Fig.~\ref{fig:react_p}. Since the proton reaction cross section is determined by inelastic processes, it is sensitive to the imaginary part of the self-energy, and is therefore vital to constraining the parameters associated with the imaginary part of the self-energy (see App.~\ref{app:params}). The neutron total cross section, which includes both elastic and inelastic processes, is shown in Fig.~\ref{fig:total_n}. The total cross section is sensitive to both the real and imaginary components of the DOM self-energy, so the good quantitative agreement over a large energy range in Fig.~\ref{fig:total_n} indicates that the positive-energy neutron parameters are well-constrained. 

Because of the dispersion relation (Eq.~\eqref{eq:dispersion}) these scattering cross sections over a wide range of positive energies also constrain the real part of the self-energy in the negative-energy domain.
The real, energy-independent component of the DOM self-energy largely determines the bound quasihole energy levels near the Fermi energy. 
The ground-state observables that are presented next receive indirect contributions from the dispersive real part that is constrained by the scattering data in~\cref{fig:xsec_p_n,fig:react_p,fig:total_n}.

\begin{figure}[b]
    \includegraphics[width=\columnwidth]{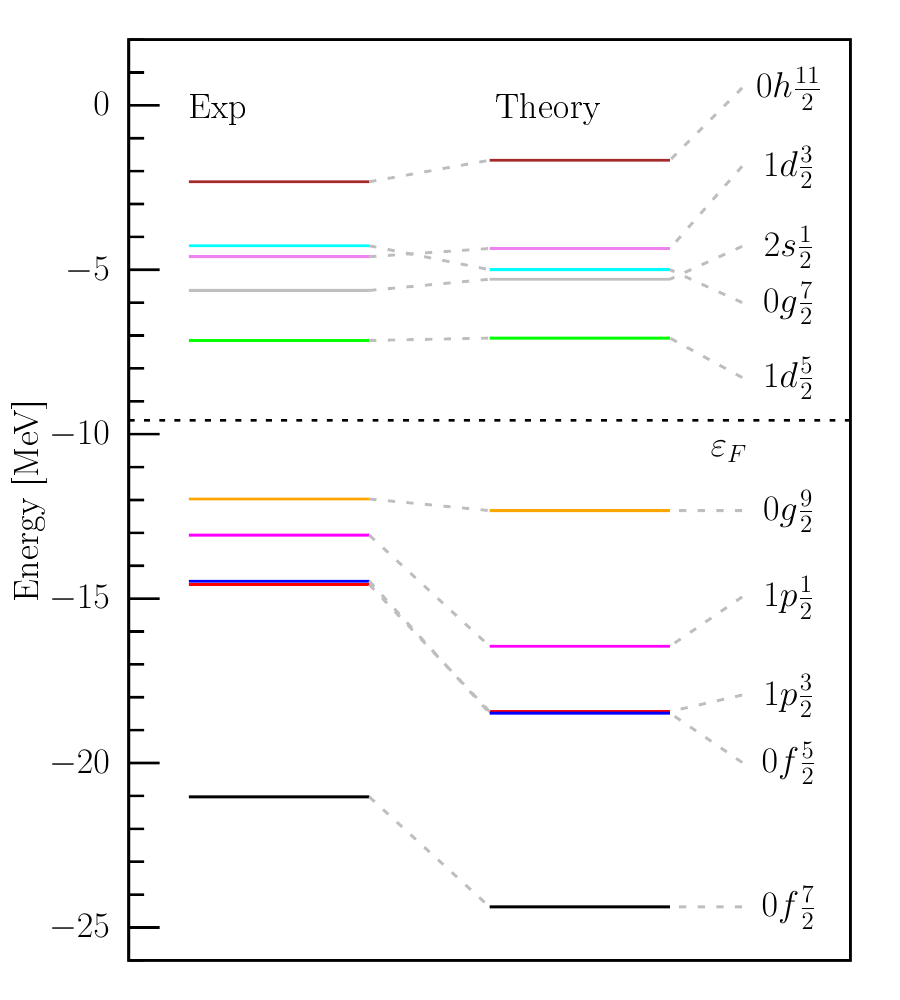}
    \caption{Neutron energy levels in $^{90}$Zr. Experimental and DOM calculated energy levels are shown on the left and right columns, respectively.}
    \label{fig:N_levels}
\end{figure}

While the protons in $^{90}$Zr form an open-shell configuration and require a pairing treatment, the neutrons ($N=50$) form a closed shell, therefore the energy levels that are generated by Eq.~\eqref{eq:schrodingerq} can immediately be compared with experimental data in Fig.~\ref{fig:N_levels}.
We only display levels near $\varepsilon_F$ in Fig.~\ref{fig:N_levels} as these are unambiguously single-particle levels (see Sec.~\ref{sec:theory}). The DOM self-energy adequately describes both quasihole ($E<\varepsilon_F$) and quasiparticle ($E>\varepsilon_F$) neutron levels noting that the DOM levels below the $0g\frac{9}{2}$ orbit represent already fragmented distributions. 

\begin{figure}[t]
    \includegraphics[width=\columnwidth]{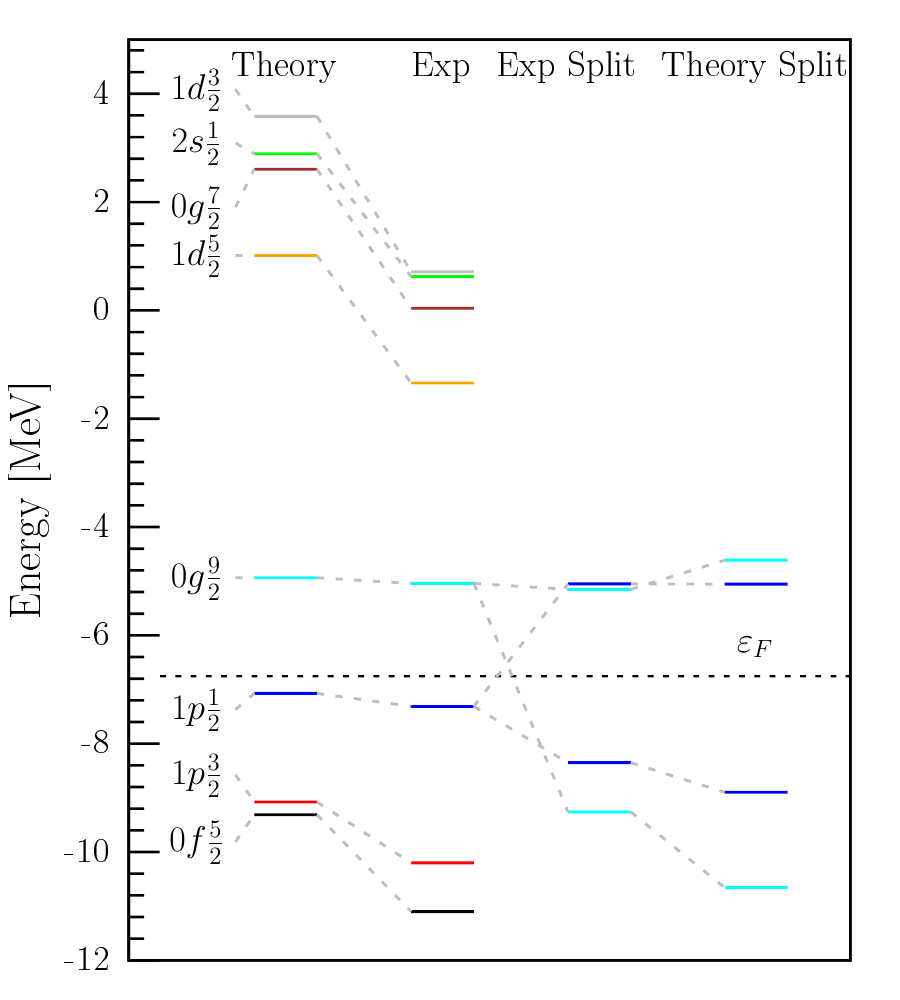}
    \caption{Proton energy levels in $^{90}$Zr. The first column is calculated employing the DOM self-energy without the pairing contribution. The second column identifies the derived experimental values that should be compared to the first column as discussed in the text. The third column identifies the experimental location of the 1$p\frac{1}{2}$ and 0$g\frac{9}{2}$ fragments above and below the Fermi energy $\varepsilon_F$. The last column shows the splitting of these levels from their position in the second column when the pairing contribution is included. The horizontal dashed line indicates the Fermi energy.}
   \label{fig:P_levels}
\end{figure}

For an open-shell proton system, ${}^{90}$Zr in this case, the proton single-particle orbits near the Fermi energy split their strength above and below the Fermi energy, as shown in the third column in Fig.~\ref{fig:P_levels} which directly identifies the experimental location of the 1$p\frac{1}{2}$ and 0$g\frac{9}{2}$ fragments. 
The experimental proton single-particle levels in the second column of Fig.~\ref{fig:P_levels} were extracted from the stripping reaction, $^{90}$Zr$(^{3}$He,$d)^{91}$Nb~\cite{knopfle1970reaction}, for levels above the Fermi energy.
The energies of each observed fragment weighted by its reported spectroscopic factor were summed for each $\ell j$ combination, and the total was divided by the sum of the spectroscopic factors in the usual way.
This yields the results shown in the second column for the 1$d\frac{5}{2}$, 0$g\frac{7}{2}$, 2$s\frac{1}{2}$, and 1$d\frac{3}{2}$ levels.
The DOM results for these levels are shown in the first column. We note that the continuum is close to the observed fragments and the location of the calculated DOM levels reflects that their spectral distribution is already substantially fragmented. This close proximity to the continuum most likely implies that there is considerable unobserved strength in the continuum.
A similar procedure was employed for the 1$p\frac{3}{2}$ and 0$f\frac{5}{2}$ strength observed in the electron-induced proton knockout reaction, $^{90}$Zr($e,e'p$)$^{89}$Y~\cite{den1988single}. No obvious strength is observed for these orbits above the Fermi energy but they are also included in the pairing treatment. The DOM results for these orbits show reasonable agreement with the corresponding experimental ones in the second column.

To describe the location of the 1$p\frac{1}{2}$ and 0$g\frac{9}{2}$ orbits in the second column of Fig.~\ref{fig:P_levels} a similar weighting was employed where Eqs.~\eqref{eq : E_plus1_lambda} and \eqref{eq : E_minus1_lambda} were utilized to extract the corresponding ``non-split" single-particle energies, $\varepsilon^{n}_{lj}$, assuming a gap parameter of 1 MeV.
The actual experimental fragmentation of these two orbits is shown in the third column. The fourth column contains the splitting of the corresponding DOM levels using the 1 MeV gap parameter for the inclusion of pairing. 
Equations~\eqref{eq:zlower} and \eqref{eq:zupper} were used to calculate the corresponding spectroscopic factors for the fragments above and below the Fermi energy.
The good agreement between the final DOM results and the experimental levels is further tested by the calculation of the $(e,e'p)$ cross section in Sec.~\ref{sec:eep}. 

\begin{figure}[t]
    \includegraphics[width=\columnwidth]{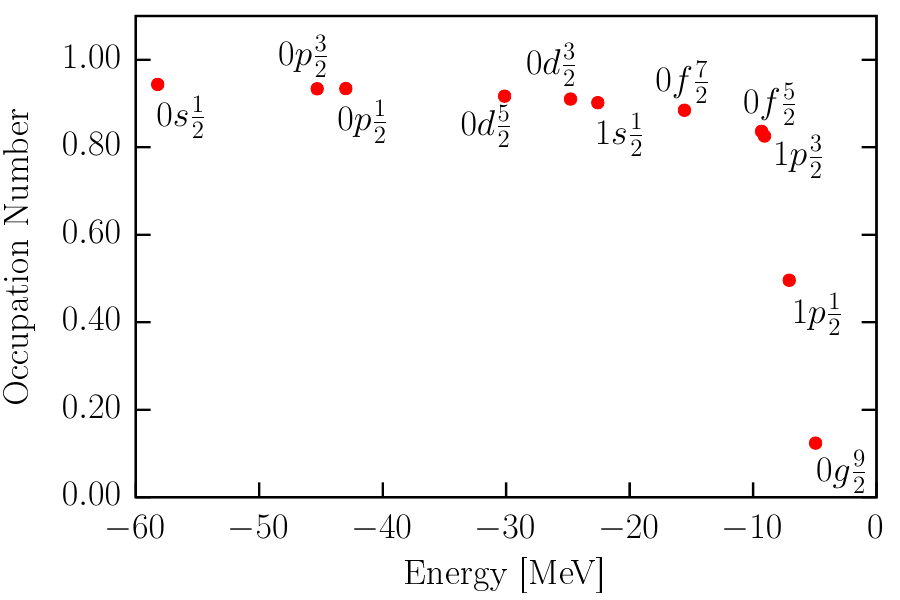}
    \caption{Proton occupation numbers for the semi-magic nucleus $^{90}$Zr. The peak positions shown in Fig.~\ref{fig:spectral_p} are labeled here with the corresponding quantum numbers. The transition from mostly occupied to mostly empty orbits is indicative of a system with pairing correlations. The spectroscopic strength of the 1$p\frac{1}{2}$ orbit near the Fermi energy is split roughly in half and together with the strength in the continuum at more negative energies its occupation number amounts to 0.50.
   \label{fig:occu}}
\end{figure}

The semi-magic nature of ${}^{90}$Zr can be illustrated by calculating the wave functions associated with the peaks of the spectral distribution in Fig.~\ref{fig:spectral_p} according to Eq.~\eqref{eq:schrodingerq}.
These wave functions $\phi^{n-}_{\ell j}(r)$, normalized to 1, can be double-folded with the hole spectral density in Eq.~\eqref{eq:spec} according to
\begin{eqnarray}
\!\!\!\!\!\!\! S_{\ell j}^{n-}(E) 
= \!\! \int \!\! dr r^2 \!\! \int \!\! dr' r'^2 \phi^{n-}_{\ell j}(r) S_{\ell j}^{h}(r ,r' ; E) \phi^{n-}_{\ell j}(r') ,
\label{eq:spechr}
\end{eqnarray}
to obtain the spectral distribution of the state associated with the chosen peak as a function of energy~\cite{Dussan:2014}.
An important sum rule is valid for the sum of the occupation number for the orbit $n_{n \ell j}$ and its depletion number $d_{n \ell j}$~\cite{Exposed!}
\begin{eqnarray}
\!\! 1 =  n_{n \ell j} + d_{n \ell j} \!\!
\label{eq:sumr} 
 =\!\! \int_{-\infty}^{\varepsilon_F} \!\!\!\! dE\ S_{\ell j}^{n-}(E) \!+ \!\! \int_{\varepsilon_F}^{\infty} \!\!\!\! dE\ S_{\ell j}^{n+}(E)  ,
\end{eqnarray}
where a corresponding integral above the Fermi energy involving the particle spectral function yields the depletion number. The sum rule is 
equivalent to $a^\dagger_{n \ell j} a_{n \ell j} +a_{n \ell j}a^\dagger_{n \ell j} =1$.
The occupation numbers associated with the peaks shown in Fig.~\ref{fig:spectral_p} are displayed in Fig.~\ref{fig:occu}.

The occupation of the deeply bound peaks reaches values slightly above 90\%, similar to the results for ${}^{40}$Ca presented in Ref.~\cite{Dussan:2014}.
For the orbits near the Fermi energy, a smooth transition of the occupation numbers is obtained, which is a characteristic feature of pairing correlations that illustrates the open-shell nature of the protons in ${}^{90}$Zr.
In paired systems, this transition is characterized by the relevant gap parameter, here 1 MeV, which means that it occurs quite rapidly.
In a closed proton-shell system like ${}^{40}$Ca, the transition from mostly occupied to mostly empty is considerably more abrupt, characteristic of a normal Fermi system~\cite{Dussan:2014}.
The 1$p\frac{1}{2}$ level near the Fermi energy obtained from Eq.~\eqref{eq:schrodingerq} has a spectroscopic strength of 0.74 before the pairing contribution is executed. 
Using Eqs.~\eqref{eq:zlower} and \eqref{eq:zupper}, the fragment below the Fermi energy acquires 0.40 and the one above 0.34.
Together with the strength of this orbit distributed at lower energy, the occupation number then becomes 0.50 and illustrates the difference of this proton system with a closed-shell one.

\begin{figure}[t]
    \includegraphics[width=\columnwidth]{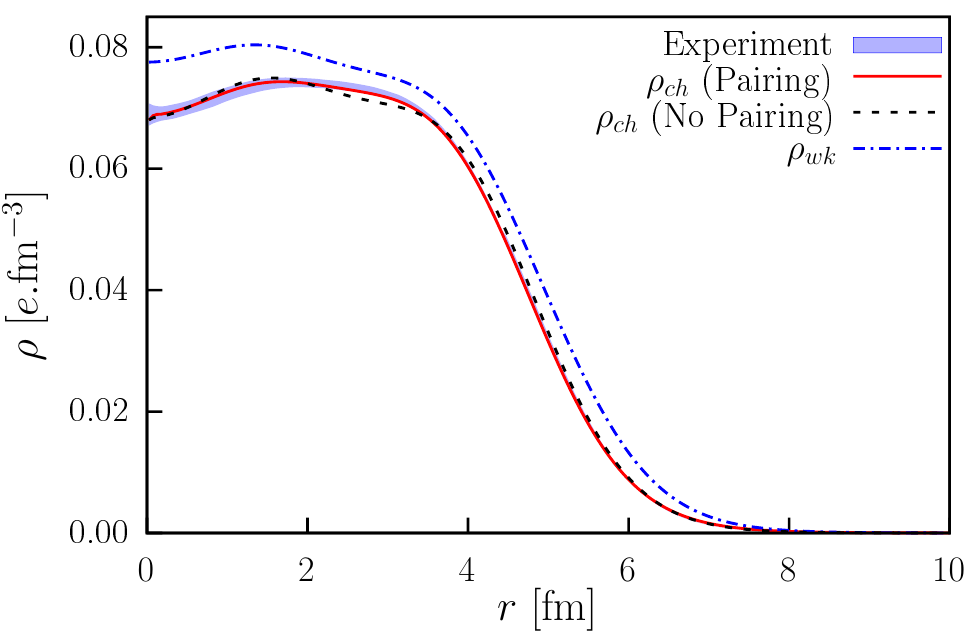}
    \caption{Experimental charge density of $^{90}$ Zr compared to DOM results. The solid line is obtained by including the pairing effect that properly splits the levels and corresponding spectroscopic factors according to the section \ref{sec:pairing}. The dashed line represents the DOM result without the pairing treatment. The dash-dotted line represents the weak charge calculated with the DOM. The experimental band represents the extracted charge density from elastic electron scattering experiments~\cite{de1987nuclear} with the appropriate error~\cite{frois1985}.}
   \label{fig:chd}
\end{figure}
The reproduction of the experimental proton levels in Fig.~\ref{fig:P_levels} including pairing is also essential to properly describe the nuclear charge density of $^{90}$Zr. The nuclear charge density, shown in Fig.~\ref{fig:chd}, is calculated using the proton and neutron point distributions from Eq.~\eqref{eq:charge}, folded with the charge distributions of the protons and neutrons, and including a spin-orbit correction (see Refs.~\cite{Brown_1979,Atkinson20} for explicit details). 
Figure~\ref{fig:chd} illustrates the DOM charge density without pairing and the corresponding improvement when it is included.
The pairing correction has succeeded in reducing the contribution of the 1$p\frac{1}{2}$ level while adding a small one from the 0$g\frac{9}{2}$ orbit leading to an accurate description of the charge density.
The weak charge density is also included in Fig.~\ref{fig:chd}. 
The resulting neutron skin corresponds to 0.078 fm with an estimated error of 50\% according to Ref.~\cite{atkinson2020dispersive}.
A discussion of this result together with updated results for ${}^{40,48}$Ca, a new result for ${}^{54}$Fe, and the earlier result for ${}^{208}$Pb is presented in Ref.~\cite{Ramon:2026}.

\begin{figure}[b]
    \includegraphics[width=\columnwidth]{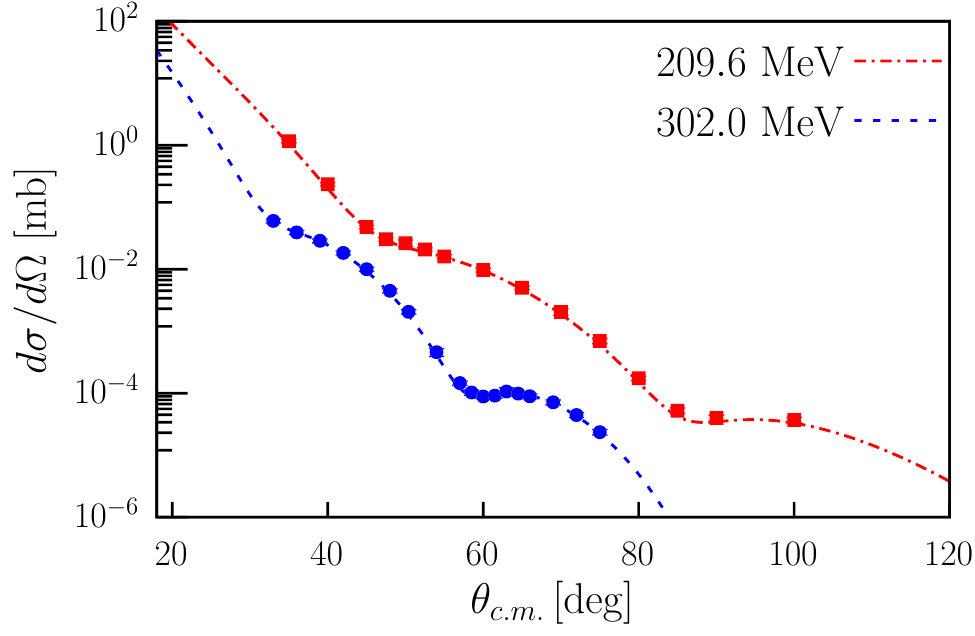}
      \caption{Differential cross section for elastic electron scattering from $^{90}$Zr. The DOM calculated charge density was used to calculate these results for 209.5 MeV and 302 MeV incident electrons. Experimental data are from Ref.~\cite{ho1972elastic}.} 
   \label{fig:E_DCS}
\end{figure} 
The quality of the charge density distribution shown in Fig.~\ref{fig:chd} is further validated by the agreement with elastic electron-scattering differential cross sections at incident electron energies of 209.6 and 302.0 MeV, calculated using the code of Ref.~\cite{salvat:2005}, as shown in Fig.~\ref{fig:E_DCS}. 
While other descriptions, mostly involving mean-field approaches~\cite{roca2008theoretical}, have difficulties reproducing the charge density near the origin for $^{90}$Zr, the DOM has successfully reproduced the experimental charge density through the full radial range. This extends previous demonstrations of the DOM's success in the doubly-magic nuclei $^{40}$Ca~\cite{Atkinson:2018}, $^{48}$Ca~\cite{calleya2025investigating}, $^{208}$Pb~\cite{atkinson2020dispersive} to the semi-magic nucleus $^{90}$Zr.

\begin{table}[t]
\caption{Integrated ground-state quantities of $^{90}$Zr. The experimental value for the charge radius, $R_\mathrm{ch}^\mathrm{exp}$, is taken from Ref.~\cite{Angeli:2013}. The experimental value for the ground-state binding energy, $E_0$, is taken from Ref.~\cite{wang2021ame}.}
\label{tab:table_pn}
\begin{ruledtabular}
\begin{tabular}{ccc}
 &
\textrm{DOM}&
\textrm{Experiment}\\
\colrule
 \vspace{0.1cm}
$N$& 50.06 & 50\\
 \vspace{0.1cm}
$Z$ & 40.03 & 40\\
 \vspace{0.1cm}
$R_\mathrm{ch}$ [fm] & 4.280 & 4.2694~$\pm$ 0.001\\
$E_0$ [MeV] & -774.34 & -783.90 \\
\end{tabular}
\end{ruledtabular}
\end{table}
With the reproduction of the experimental charge density, the DOM naturally describes the root-mean-square (RMS) charge radius, $R_\mathrm{ch}$, in good agreement with the experimental value, as seen in Table.~\ref{tab:table_pn}. Furthermore, the number of neutrons and protons determined as the $0^\mathrm{th}$ moments of the neutron and proton distributions in Eq.~\eqref{eq:charge}, respectively, are consistent (see Table.~\ref{tab:table_pn}). 
The experimental ground-state energy, $E_0$, of $^{90}$Zr is also reproduced within 1.5\% by employing the Migdal-Galitski sum rule with the DOM spectral functions~\cite{Exposed!,Atkinson2020B}. The uncertainty in the pairing gap (see Sec.~\ref{sec:eep}) propagates to the calculated binding energy to an insignificant amount of 0.1\%.

\section{$^{90}$\textmd{Z\NoCaseChange{r}}$(e,e'p)$$^{89}$\textmd{Y} reaction}
\label{sec:eep}

The DOM provides all the ingredients for the calculation of the DWIA description of the $(e,e'p)$ reaction.
For $^{40}$Ca it was confirmed that the DOM overlap function, normalized with the corresponding spectroscopic factor, together with the appropriate DOM distorted wave provides an excellent description of the cross sections for the removal of the valence protons~\cite{Atkinson:2018}.
The same study also confirmed that parallel kinematics chosen in the Nikhef experiments together with energies of the outgoing protons around 100 MeV allow an accurate description employing the DWIA.
Corresponding results for ${}^{48}$Ca confirm this conclusion~\cite{Atkinson:2019}.

In the DWIA, the final-state interaction between the ejected proton and the residual nucleus is incorporated by generating a distorted wave determined by the optical potential at the relevant proton energy.  
The cross section is calculated from the hadron tensor, $W^{\mu\nu}$,  which contains matrix elements of the nuclear charge-current density, $J^\mu$~\cite{ElectroResponse}.  
Using the DWIA, which assumes that the virtual photon exchanged by the electron couples to the 
same proton that is detected 
~\cite{Giusti:1988,Boffi:1980}, the nuclear current can be written as
\begin{equation}
      J^\mu(\bm{q}) = \int d\bm{r}e^{i\bm{q}\cdot\bm{r}}\chi^{(-)*}_{E\ell j}(\bm{r})(\hat{J}^\mu_{\text{eff}})_{E\ell j}(\bm{r})\psi^n_{\ell j}(\bm{r})\sqrt{\mathcal{Z}^n_{\ell j}},
      \label{eq:current}
\end{equation}
   where $\chi^{(-)*}_{E}(\bm{r})$ is the outgoing proton distorted wave~\cite{ElectroResponse}, 
   $\psi^n_{\ell j}$ is the overlap function, $\mathcal{Z}^n_{\ell j}$ its normalization, $\bm{q} = \bm{k_f} - \bm{k_i}$ is the electron three-momentum transfer, and 
   $\hat{J}^\mu_{\text{eff}}$ is the effective current operator~\cite{ElectroResponse}. 
   The incoming and outgoing electron waves
   are treated within the effective momentum approximation, where the waves are represented by plane waves with effective momenta to account for distortion from the interaction with the target 
   nucleus~\cite{Giusti:1987}
   \begin{equation}
      k_{i(f)}^{\text{eff}} = k_{i(f)} + \int d\bm{r}V_c(\bm{r})\phi_{\ell j}^2(\bm{r}),
      \label{eq:effective}
   \end{equation}
   where $V_c(\bm{r})$ is the Coulomb interaction.
   This alters Eq.~(\ref{eq:current}) by replacing $\bm{q}$ with $\bm{q}_{\text{eff}}$.

In the plane-wave impulse approximation (PWIA), in which the outgoing proton wave is approximated by a plane wave, the $(e,e'p)$ cross section can be factorized into an off-shell electron-proton cross section, $\sigma_{ep}$, approximated from the on-shell one using the $\sigma_{\text{cc1}}$ model as proposed in~\cite{deForest:1983} and the spectral function~\cite{ElectroResponse}.
This separation does not hold true for the DWIA, but the displayed cross sections, both the experimental and theoretical ones, have been divided by the $\sigma_{\text{cc1}}$ cross section,
\begin{equation}
    S(E_m,p_m,p')=
\frac{1}{K\sigma_{ep}}
\frac{d^{6}\sigma}
{dk'_0\, d\Omega_{k'}\, dp'_0\, d\Omega_{p'}} 
\end{equation}
and the momentum-density distribution $\rho_\alpha(p_m)$
\begin{equation}
    \rho_{\alpha}(p_m)
=
\int_{\Delta E_m}
S(E_m,p_m,p')\, dE_m .
\end{equation}
The distortion of the proton waves is calculated using a partial-wave expansion.
The momentum distributions are calculated by adapting a recent version of the DWEEPY code~\cite{Giusti:2011} to utilize the DOM overlap functions with their associated normalization, and the DOM distorted waves~\cite{Atkinson:2018}.

\begin{figure}[b]
    \includegraphics[width=\columnwidth]{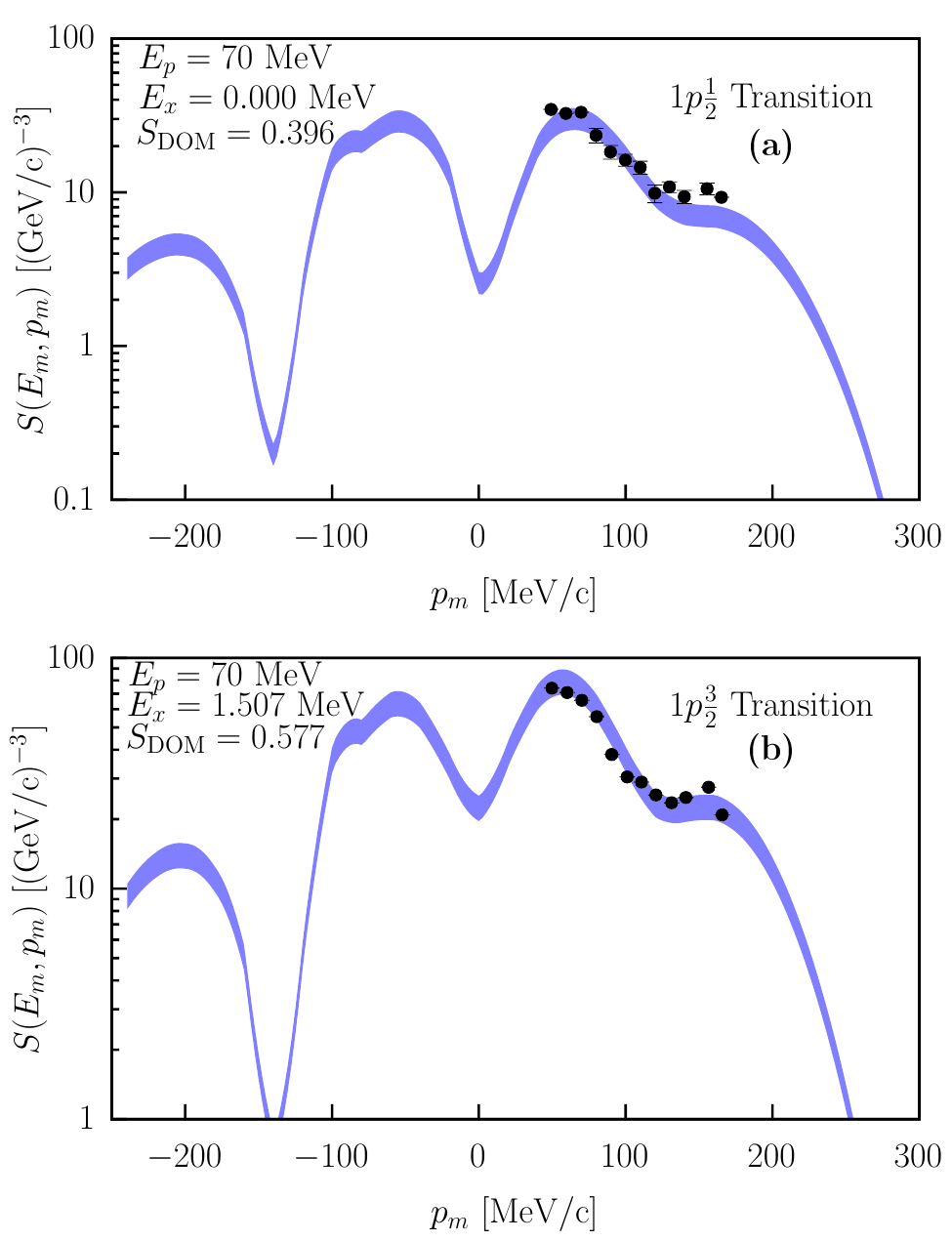}
      \caption{$^{90}$Zr$(e,e'p)$$^{89}$Y momentum densities in parallel kinematics at an outgoing proton kinetic energy of $70$~MeV for $1p\frac{1}{2}$ (a) and $1p\frac{3}{2}$ (b). Shaded bands represent results incorporating DOM ingredients. The width of the uncertainty is related to the value of the pairing gap and the approximation of the Coulomb distortion as discussed in the text. 
      }
   \label{fig:eep_p12_p32_70}
\end{figure}

\begin{figure}[h]
            \includegraphics[width=\columnwidth]{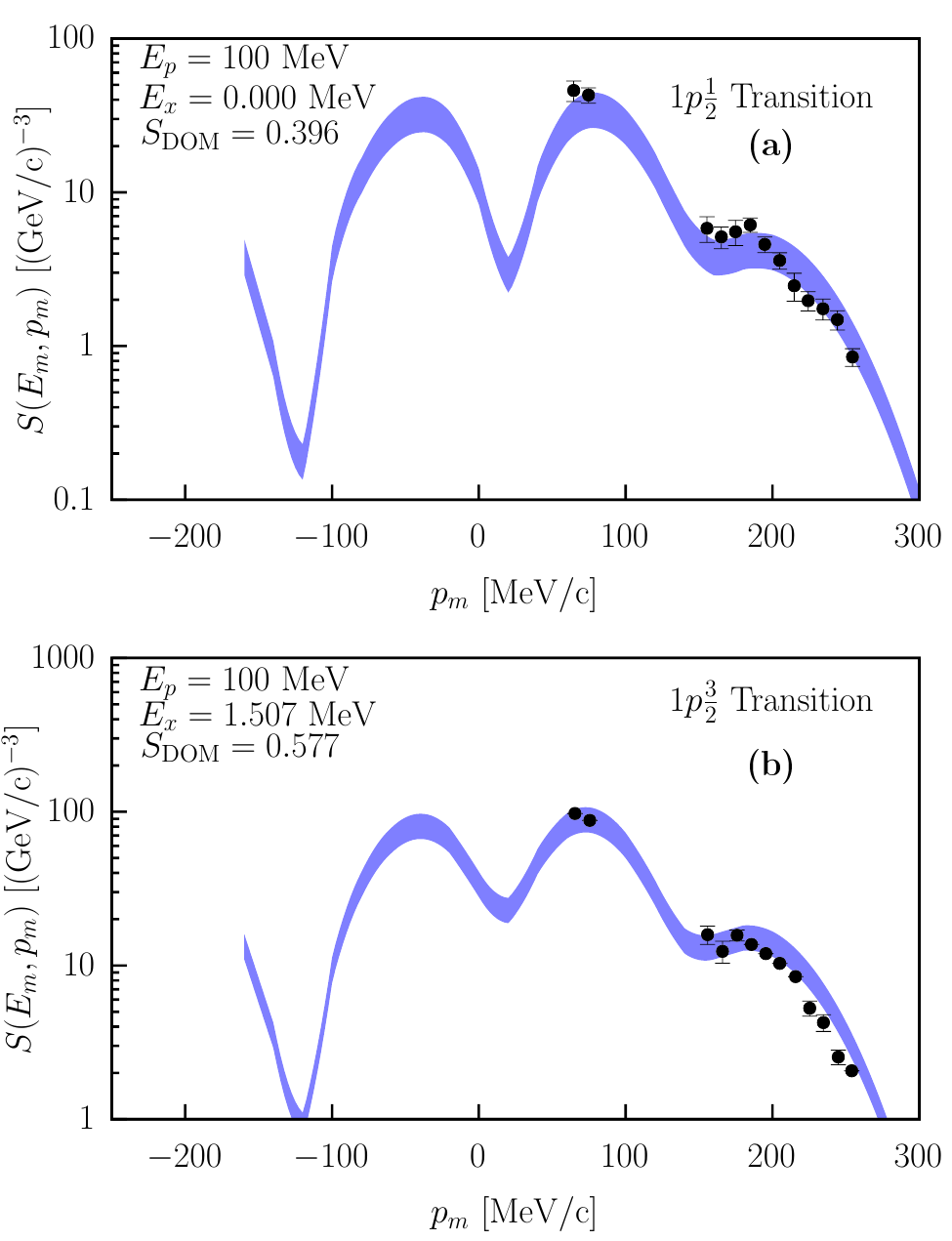}
      \caption{As for Fig.~\ref{fig:eep_p12_p32_70} but for an outgoing proton energy of $100$~MeV.}
   \label{fig:eep_p12_p32_100}
\end{figure}

The $^{90}$Zr$(e,e'p)^{89}$Y experiments reported in Ref.~\cite{Denherder:1988} were among the very first that employed a high-resolution coincidence set-up.
Results in Ref.~\cite{Denherder:1988} have been reported for 70 and 100 MeV outgoing protons covering the discrete transitions for 1$p\frac{1}{2}$, 0$g\frac{9}{2}$, 1$p\frac{3}{2}$, and 0$f\frac{5}{2}$ fragments with the latter two orbits exhibiting two distinct contributions.
The success of the DOM also clarifies some of its limitations. In particular, it is not (yet) possible to describe the details of the strength fragmentation near the Fermi energy except by introducing a smoothly increasing imaginary part of the self-energy away from the Fermi energy.
This is less important for the ground-state transition in the $Z$-1 system as typically only a single fragment is observed.
Therefore, the transition to the $\frac{1}{2}^-$ ground state of ${}^{89}$Y only requires the split of the spectral strength of the $1p\frac{1}{2}$ spectroscopic factor due to pairing as discussed in Sec.~\ref{sec:pairing}. 
The resulting strength of 0.396 is compared to the extracted Nikhef result of 0.360 in Table~\ref{tab:table1}.
\begin{table}[tb]
\caption{Comparison of spectroscopic factors extracted by the Nikhef analysis~\cite{Denherder:1988} using the local Schwandt optical potential~\cite{Schwandt:1982} to the normalization of the corresponding overlap functions obtained in the present analysis from the DOM.}
\label{tab:table1}
\begin{ruledtabular}
\begin{tabular}{ccc}
\textrm{${n}{\ell}{j}$} & \multicolumn{2}{c}{\textrm{\textrm{$\mathcal{Z}^{n}_{\ell j}$}}}\\
 \noalign{\vskip 0.1cm}
 \cline{2-3}
 \noalign{\vskip 0.1cm}
 & \textrm{DOM} & \textrm{Ref.}~\cite{Denherder:1988}\\
\colrule
 \noalign{\vskip 0.1cm}
 \vspace{0.1cm}
$1p{\frac{1}{2}}$& 0.396 & 0.360\\
 \vspace{0.1cm}
$1p{\frac{3}{2}}$ & 0.577 & 0.465\\
 \vspace{0.1cm}
$0f{\frac{5}{2}}$ & 0.522 & 0.462\\
 \vspace{0.1cm}
$0g{\frac{9}{2}}$ & 0.047 & 0.054\\
\end{tabular}
\end{ruledtabular}
\end{table}
The calculated DOM cross sections for this transition are compared with data in Figs.~\ref{fig:eep_p12_p32_70}(a) and \ref{fig:eep_p12_p32_100}(a) for outgoing proton energies of 70 and 100 MeV, respectively.
Previous results for DOM generated spectroscopic factors in Refs.~\cite{Atkinson:2018,Atkinson:2019} demonstrated that the experimental errors imply an uncertainty of about $\pm 0.05$ for large fragments. 
The uncertainty in Refs.~\cite{Atkinson:2018,Atkinson:2019} was found through a bootstrap method.
Modified data sets were created from the experimental data used to constrain the DOM self-energy by randomly renormalizing each angular distribution or excitation function within the experimental error to incorporate fluctuations from the systematic errors.
Since the experimental coverage of the relevant elastic-scattering and structure data in $^{90}$Zr is comparable to that of $^{40}$Ca and $^{48}$Ca, we assume a similar result holds here with a similar relative error for smaller fragments.
An alternative statistical approach employing the DOM was introduced in Refs.~\cite{Pruitt:2020,Pruitt:2020C} which may provide further insights into error estimates in the future.

\begin{figure}[t]
            \includegraphics[width=\columnwidth]{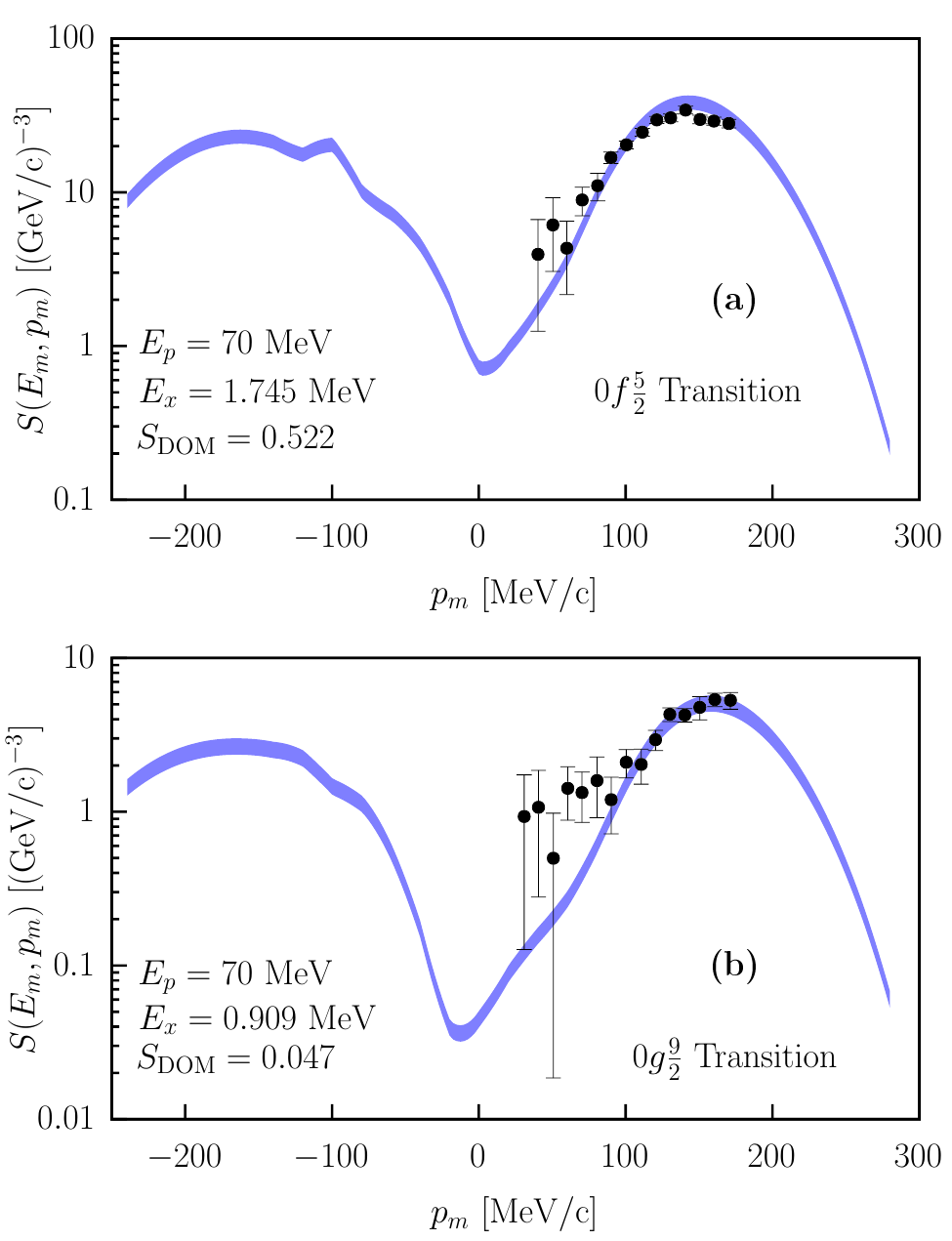}
      \caption{$^{90}$Zr$(e,e'p)$$^{89}$Y cross sections in parallel kinematics at an outgoing proton kinetic energy of $70$~MeV for the main $0f\frac{5}{2}$ transition at 1.745 MeV in (a). Shaded bands represent results from the DOM ingredients as discussed in the text. In part (b) the transition involving the $0g\frac{9}{2}$ orbit,  at 0.909 MeV in $^{89}$Y is displayed. }
      \label{fig:eep_f52_g92_70}
\end{figure} 
\begin{figure}[t]
    \includegraphics[width=\columnwidth]{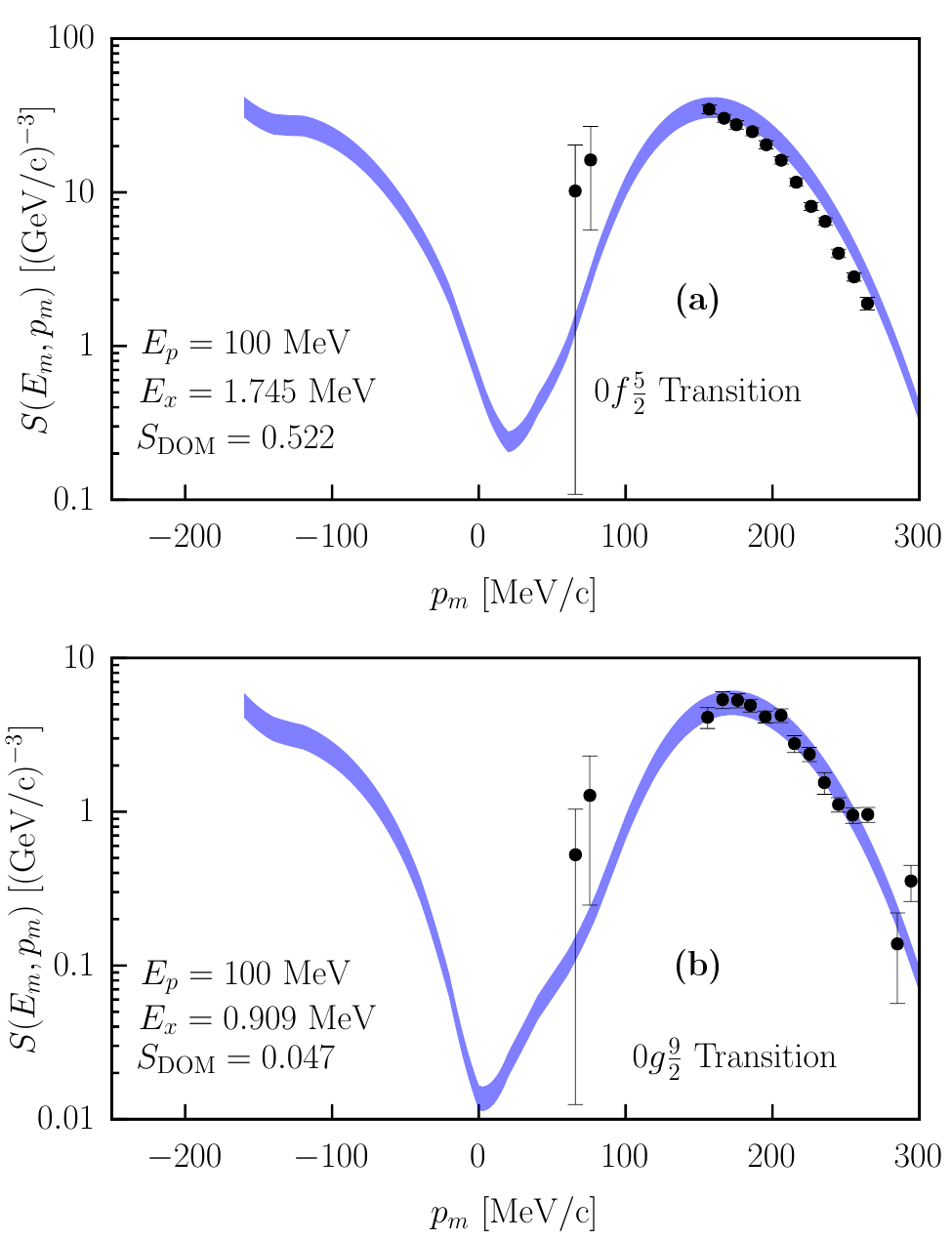}
    \caption{As for Fig.~\ref{fig:eep_f52_g92_70} but for an  outgoing proton energy of $100$~MeV.}
   \label{fig:eep_f52_g92_100}
\end{figure} 

The bands in \cref{fig:eep_p12_p32_70,fig:eep_p12_p32_100,fig:eep_f52_g92_70,fig:eep_f52_g92_100} represent a combination of the 5\% statistical uncertainty discussed above and the two sources of systematic uncertainty described below.
The adopted pairing gap of $\Delta = 1$ MeV is not unique. As an alternative, it may be estimated using 
\textit{e.g.} Eq.~\eqref{eq:E_pairing}. To account for this ambiguity, we varied the pairing gap according to  
$\Delta = 1 \pm0.1169$~MeV when evaluating the 
$^{90}$Zr$(e,e'p)^{89}$Y cross sections.
A second source of systematic uncertainty arises from the treatment of Coulomb distortion of the electron wave functions. In the Nikhef analysis, the effect of Coulomb distortion was calculated using the ``high-energy approximation''~\cite{Denherder:1988,Herder1987Single}. The Coulomb potential enhances the momentum transfer and causes an increased electron flux near the nucleus because of the larger electron energy in the Coulomb potential. According to this approximation, the electron energies are enhanced by $f(Z\alpha/R_C)$, where $Z$ is the number of protons, $\alpha$ is the fine-structure constant, and $R_C$ is the Coulomb radius. 
Within this approximation, the correction both shifts the momentum distributions along the momentum axis and produces a momentum-dependent enhancement of the cross sections. 
The pre-factor $f$ is typically varied between $ 1.10$ and $1.50$ depending on the orbital. The DOM calculations were therefore varied between these limits.
Taking these statistical and systematic uncertainties into account, the agreement between the calculated and measured momentum distributions for $1p\frac{1}{2}$ removal in Figs.~\ref{fig:eep_p12_p32_70}(a) and \ref{fig:eep_p12_p32_100}(a) is satisfactory, given that this represents the first DOM treatment of an open-shell proton system. 

The fragmentation of more deeply bound orbits must be addressed before a comparison with data can be made as documented in Ref.~\cite{Atkinson:2018}.
If available, the remedy is to incorporate experimental information of this fragmentation that is documented for the $(e,e'p)$ reaction for both ${}^{40}$Ca and ${}^{48}$Ca in the experimental analysis.
Such a procedure is not sufficient for the $^{90}$Zr$(e,e'p)^{89}$Y reaction as only one additional discrete transition for both $1p\frac{3}{2}$ and $0f\frac{5}{2}$ is identified in Ref.~\cite{Denherder:1988}.

The experimental information of fragmentation observed in the description of the $^{90}$Zr$(e,e'p)^{89}$Y reaction is sparse.
These additional discrete poles are not explicitly included in the DOM, although there is a smooth energy-dependent imaginary term in the self-energy to approximate their effect on the spectral strength~\cite{Exposed!}. 
The calculated DOM spectroscopic factor must be reduced by the experimental strength observed in the neighborhood of the quasihole energy.
This effect is incorporated for the 1$p\frac{3}{2}$, and 0$f\frac{5}{2}$ orbits by enforcing that the ratio between the strength of the peak to the total spectral strength shown in the energy domain of Fig.~(\ref{fig:spectral_p}) is the same between the data as for the DOM, 
   \begin{equation}
      \frac{\mathcal{Z}_F^{\text{DOM}}}{\int dE\ S^{\text{DOM}}(E)} = \frac{\mathcal{Z}_F^{\text{exp}}}{\int dE\ S^{\text{exp}}(E)}.
      \label{eq:ratio}
   \end{equation}
Accounting for the contributions to the momentum distribution from different energies by scaling the DOM spectroscopic factor is further justified by observing that the shape of the momentum distribution calculated at similar energies is identical, with the strength being the only difference~\cite{Denherder:1988} and also documented in Ref.~\cite{Kramer:1989}. 
Table 10 of Ref.~\cite{Denherder:1988} identifies two fragments associated with $1p\frac{3}{2}$ removal carrying strengths of 0.465 and 0.030, respectively.
For the $0f\frac{5}{2}$ removal this table reports two fragments corresponding to 0.462 and 0.048, respectively.
The larger fragments are also listed in Table~\ref{tab:table1}.
The DOM entry for these two fragments corresponds to 0.577 and 0.522, respectively, and includes this experimental correction just described, the theoretical correction discussed below, 
as well as a correction due to the inclusion of pairing correlations.
The latter amounts to only a few \% in this case.

The DOM spectral strength for the $1p\frac{3}{2}$ and $0f\frac{5}{2}$ orbits shown in Fig.~\ref{fig:spectral_p} contains additional strength in the interval from -20 MeV to the Fermi energy compared to the spectroscopic factors obtained from Eq.~\eqref{eq:schrodingerq}.
The value of -20 MeV corresponds to the minimum of the $p\frac{3}{2}$ strength.
In this interval, the character of the spectral density is dominated by the corresponding solutions of Eq.~\eqref{eq:schrodingerq}.
The strength integrated in this interval is 0.797 for the $1p\frac{3}{2}$ and 0.757 for the $0f\frac{5}{2}$ orbit compared to the normalization of the solutions of Eq.~\eqref{eq:schrodingerq} of 0.711 and 0.670, respectively.
We assume that this extra theoretical strength describes additional fragmentation of the strength not yet reported experimentally.
When comparing with the experimental $(e,e'p)$ cross sections we have noted that correcting the spectroscopic factor from Eq.~\eqref{eq:spec} by the ratio of these numbers in addition to the experimental correction discussed above, yields sensible results.

The resulting cross section for the main $1p\frac{3}{2}$ fragment is shown in Fig.~\ref{fig:eep_p12_p32_70}(b) and \ref{fig:eep_p12_p32_100}(b), at 70 and 100 MeV proton energy, respectively.
Agreement with the data is similarly satisfactory as for the ground-state transition.
The results for the main $0f\frac{5}{2}$ fragment are displayed for the 70 and 100 MeV cases in Figs.~\ref{fig:eep_f52_g92_70}(a) and \ref{fig:eep_f52_g92_100}(a) with similar good agreement.

We note that the spectroscopic strength for these transitions in the DOM analysis is higher than that extracted from the Nikhef analysis, as compared in Table~\ref{tab:table1}.
A similar difference was documented in Refs.~\cite{Atkinson:2018,Atkinson:2019} and is most likely due to the use of a nonlocal potential in the DOM case.
It is also noteworthy that the agreement at the different proton energies is similar to that of the ${}^{40}$Ca case in Ref.~\cite{Atkinson:2018}.

The small fragment involving the $0g\frac{9}{2}$ level occurs below the Fermi energy as a consequence of the pairing treatment shown in Fig.~\ref{fig:P_levels}.
The corresponding strength is only 0.023.
We also include the strength found by integrated the $0g\frac{9}{2}$ spectral function from -20 MeV to the Fermi energy (similar to the procedure for the $1p\frac{3}{2}$ and $0f\frac{5}{2}$ orbits), resulting in a final result of 0.047 as shown in Table~\ref{tab:table1} and employed in the calculation for this transition in Figs.~\ref{fig:eep_f52_g92_70}(b) and \ref{fig:eep_f52_g92_100}(b).

\section{Conclusion}
\label{sec:conclusion}

We have extended the nonlocal dispersive optical model to provide a complete description of the semi-magic nucleus $^{90}$Zr by incorporating proton pairing correlations.
The DOM self-energy was constrained to reproduce elastic-scattering data as well as energy levels, total binding energy, particle number, and charge density.

The 40 protons in $^{90}$Zr form an open-shell configuration 
which complicates the typical method of constraining the proton self-energy in the negative energy domain near the Fermi energy.
A two-step procedure was applied in which the conventional DOM calculation was performed as a first step with a subsequent treatment of pairing correlations following Ref.~\cite{Migdal:1967}.
A single gap parameter close to the empirical value allows for an adequate description of the fragments above and below the Fermi energy of the $1p\frac{1}{2}$ and $0g\frac{9}{2}$ single-particle levels as shown in Fig.~\ref{fig:P_levels}. This accurate fragmentation is instrumental in consistently describing all available $^{90}$Zr$(e,e'p)$$^{89}$Y momentum distributions from discrete transitions displayed in \cref{fig:eep_p12_p32_70,fig:eep_p12_p32_100,fig:eep_f52_g92_70,fig:eep_f52_g92_100}.

The inclusion of pairing also leads to an accurate description of the charge density distribution as evidenced by the quantitative agreement with both the experimental charge density in Fig.~\ref{fig:chd} and corresponding elastic electron-scattering differential cross sections in Fig.~\ref{fig:E_DCS}.
This extends the DOM's demonstrated ability to describe density distributions from the closed-shell nuclei
 $^{40}$Ca~\cite{Atkinson:2018}, $^{48}$Ca~\cite{calleya2025investigating}, and $^{208}$Pb~\cite{atkinson2020dispersive} to the semi-magic nucleus $^{90}$Zr.
 Based on this description, we predict a neutron skin of $R_{skin} = 0.078 \pm 0.039$~fm for $^{90}$Zr.

 The present work demonstrates that the DOM framework can be systematically extended beyond doubly closed-shell nuclei without sacrificing its simultaneous description of nuclear structure and reaction observables. 
 Building on the present extension of the DOM, future work can incorporate a Hartree-Fock potential based on the underlying nucleon-nucleon interaction to better describe $N\ne Z$ level structures~\cite{RevModPhys.92.015002}, utilize shell-model or microscopic inputs to study low-energy fragmentation patterns in greater detail~\cite{Brown01}, and apply coupled-channel and collective-model formalism to account for deformation effects in open-shell systems (see as an example~\cite{PhysRevC.94.064605}).
 
 This provides a path toward understanding the evolution of single-particle strength and many-body correlations across both closed- and open-shell nuclei within a unified framework.

\begin{acknowledgments}
This work was supported by the U.S. National Science Foundation under grants PHY-2207756 and PHY-2512895.
\end{acknowledgments}

\appendix

\section{Parametrization of DOM Self-Energy}
\label{sec:param}
We provide a detailed description of the parametrization of the proton and neutron self-energies in
$^{90}$Zr used in the fits to bound and scattering data. The functional forms are equivalent to
those used for the $^{48}$Ca potential, detailed in Ref.~\cite{Atkinson:2019}. Parameters
which are allowed to be different for protons and neutrons will contain $(n,p)$ notation.  Asymmetry
terms have been added to the amplitudes of many of the components in the form
$\pm V_{(p,n)}\frac{N-Z}{A}$ where here only, the $+$ refers to protons and $-$ to neutrons.
Elsewhere, $\pm$ in superscripts and subscripts refer to above ($+$) and below ($-$) the Fermi
energy, $\varepsilon_F$. 

We use a simple Gaussian nonlocality in all instances \cite{Perey:1962} and restrict the nonlocal contributions to the HF term and to the volume and surface contributions to the imaginary part of the potential.
We write the HF self-energy term in the following form
with spin-orbit and a local  Coulomb contribution.
\begin{equation}
\Sigma_{HF}(\bm{r},\bm{r}') = \Sigma^{nl}_{HF}(\bm{r},\bm{r}') + V^{nl}_{so}(\bm{r},\bm{r}') + \delta(\bm{r}-\bm{r}') V_C(r),
\end{equation}
The nonlocal term is split into a volume and a narrower Gaussian term of opposite sign to make the final potential have a wine-bottle shape.
\begin{equation}
\Sigma_{HF}^{nl}\left( \bm{r},\bm{r}' \right) = -V_{HF}^{vol}\left( \bm{r},\bm{r}'\right) + V_{HF}^{wb}(\bm{r},\bm{r}'),
\end{equation}
where the volume term is given by
\begin{align}
   &V_{HF}^{vol}\left( \bm{r},\bm{r}' \right) =  V^{HF}_{sym} ,f \left (
   \tilde{r},r^{HFsym}_{(p,n)},a^{HFsym}_{(p,n)} \right ) \label{eq:HFv}\\
   &\times\left [ \frac{1}{1+x_{sym}} H \left( \bm{s};\beta^{vol_1}_{sym} \right) + \frac{x_{sym}}{1+x_{sym}} H \left( \bm{s};\beta^{vol_2}_{sym}\right) \right ]  \nonumber \\
   &\pm V_{(p,n)}^{HFasy} \frac{N-Z}{A}f \left ( \tilde{r},r^{HFaym}_{(p,n)},a^{HFasy}_{(p,n)} \right
   )\times \nonumber \\
   &\left [ \frac{1}{1+x_{sym}} H \left( \bm{s};\beta^{vol_1asy}_{(p,n)} \right)  +
   \frac{x_{sym}}{1+x_{sym}} H \left( \bm{s};\beta^{vol_2asy}_{sym}\right) \right ].\nonumber
\end{align} 
allowing for two different nonlocalities with different weights ($0 \le x_{sym} \le1$ in
Eq.~\eqref{eq:HFv}). 
With   the notation $\tilde{r} =(r+r')/2$ and $\bm{s}=\bm{r}-\bm{r}'$,
the wine-bottle ($wb$) shape is described by
\begin{equation}
V_{HF}^{wb}(\bm{r},\bm{r}') = V^{wb}_{(p,n)}  \exp{\left(- \tilde{r}^2/(\rho^{wb}_{(p,n)})^2\right)} H \left( \bm{s};\beta^{wb} \right ),
\label{eq:wb}
\end{equation}
where the nonlocality in Eq.~\eqref{eq:wb} is represented by a Gaussian form
\begin{equation}
H \left( \bm{s}; \beta \right) = \exp \left( - \bm{s}^2 / \beta^2 \right)/ (\pi^{3/2} \beta^3) .
\nonumber
\end{equation}
As usual, we employ Woods-Saxon form factors 
\begin{align}
f(r,r_{i},a_{i})=\left[1+\exp \left({\frac{r-r_{i}A^{1/3}}{a_{i}}%
}\right)\right]^{-1} .
\label{Eq:WS}
\end{align}
The Coulomb term is obtained from the experimental charge density distribution for $^{48}$Ca~\cite{de1987nuclear}.

The local spin-orbit interaction is given by
\begin{align}
      V_{so}(\bm{r},\bm{r'})= \left( \frac{\hbar}{m_{\pi }c}\right) ^{2}
      V^{so}_{(p,n)}\frac{1}{\tilde{r}}\frac{d}{d\tilde{r}}f(\tilde{r},r^{so}_{(p,n)},a^{so})\nonumber \\
      \times\bm{\ell}\cdot \bm{\sigma}
      H(\bm{s};\beta^{so}), 
   \label{eq:HFso}
\end{align}
where $\left( \hbar /m_{\pi }c\right) ^{2}$=2.0~fm$^{2}$ 
as in Ref.~\cite{Mueller:2011}.

The fully-nonlocal imaginary part of the DOM self-energy has the following form,
\begin{align}
\label{eq:imnl}
&\textrm{Im}\ \Sigma^{nl}(\bm{r},\bm{r}';E) = \hspace{5cm} \\ \nonumber  
&-W^{vol}_{0\pm}(E) f\left(\tilde{r};r^{vol}_{\pm (p,n)};a^{vol}_{\pm}\right)H \left( \bm{s}; \beta_{\pm (p,n)}^{vol}\right) \hspace{1cm} \\ \nonumber
&+ 4a^{sur}_{\pm sym}W^{sur0}_{\pm}\left( E\right)H \left( \bm{s}; \beta_{\pm}^{sur0}\right) \frac{d}{d \tilde{r} }f(\tilde{r},r^{sur0}_{\pm(p,n))},a^{sur}_{\pm sym}) \\ \nonumber
& + 4 a^{sur}_{\pm sym} W^{sur}_{\pm}(E) H \left( \bm{s};\beta^{sur}_{\pm (p,n)} \right ) \frac{d}{d \tilde{r} }f(\tilde{r},r^{sur}_{\pm(p,n)},a^{sur}_{\pm(p,n)}) \\ \nonumber 
&+ \textrm{Im}\Sigma_{so}(\bm{r},\bm{r}';E).
\end{align}
Note that the parameters relating to the shape of the imaginary spin-orbit term
are the same as those used for the real spin-orbit term.
At energies well removed
from $\varepsilon_F$, the form of the imaginary volume potential should not be
symmetric about $\varepsilon_F$ as indicated by the $\pm$ notation in the subscripts and superscripts~\cite{Dussan:2011}.
While more symmetric about $\varepsilon_F$, we have allowed a similar option for the surface absorption that is also supported by theoretical work reported in Ref.~\cite{Waldecker:2011}.
Allowing for the aforementioned asymmetry around $\varepsilon_F$ the following form was assumed for the depth of the volume potential~\cite{Mueller:2011}

\begin{widetext} 
\begin{equation}
W^{vol}_{0\pm}(E) =  \Delta W^{\pm}_{NM}(E) +  
\begin{cases}
0 & \text{if } |E-\varepsilon_F| < \mathcal{E}^{vol}_{\pm} \\
A^{vol}_{\pm(p,n)}  \frac{\left(|E-\varepsilon_F|-\mathcal{E}^{vol}_{\pm}\right)^4}
{\left(|E-\varepsilon_F|-\mathcal{E}^{vol}_{\pm}\right)^4 + (B^{vol}_{\pm})^4} & 
 \text{if } |E-\varepsilon_F| > \mathcal{E}^{vol}_{\pm} ,
\end{cases} 
\label{eq:volumeS}
\end{equation}
\end{widetext}
where $\Delta W^{\pm}_{NM}(E)$ in Eq.~\eqref{eq:volumeS} is the energy-asymmetric correction modeled after
nuclear-matter calculations. The asymmetry above and below $\varepsilon_F$ is essential to accommodate the Jefferson Lab $(e,e'p)$ data at large missing energy.
The energy-asymmetric correction was taken as 
\begin{widetext} 
\begin{equation}
\Delta W^{\pm}_{NM}(E)=
\begin{cases}
\alpha^{sym}_{(p,n)} A^{vol}_{+(p,n)}\left[ \sqrt{E}+\frac{\left( \varepsilon_F+\mathbb{E}_{+}\right) ^{3/2}}{2E}-\frac{3}{2}
\sqrt{\varepsilon_F+\mathbb{E}_{+}}\right] & \text{for }E-\varepsilon_F>\mathbb{E}_{+} \\ 
-  A^{vol}_{-(p,n)} \frac{(\varepsilon_F-E-\mathbb{E}_{-})^2}{(\varepsilon_F-E-\mathbb{E}_{-})^2+(\mathbb{E}_{-})^2} & \text{for }E-\varepsilon_{F}<-\mathbb{E}_{-} \\ 
0 & \text{otherwise},
\end{cases} 
\label{eq:Wnmnl}
\end{equation} 
\end{widetext}
where $E$ in Eq.~\eqref{eq:Wnmnl} corresponds to the center-of-mass energy.

To describe the energy dependence of surface absorption, we employed the form of
Ref.~\cite{Charity:2007}, but include two components, one with symmetric parameters, the other with
asymmetric parameters.
\begin{align}
W^{sur0}_{\pm}\left( E\right) =\omega _{4}(E,A^{sur0}_{\pm},B^{sur0_1}_{\pm},0)- \nonumber \\
\omega_{2}(E,A^{sur0}_{\pm},B_{\pm}^{sur0_2},C^{sur0}_{\pm}),  \label{eq:paranl} 
\end{align}
\begin{align}
  W^{sur}_{\pm(p,n))}\left( E\right) =\omega _{4}(E,A^{sur}_{\pm(p,n)},B^{sur_1}_{\pm(p,n)},0)- \nonumber \\
  \omega_{2}(E,A^{sur}_{\pm(p,n)},B_{\pm(p,n)}^{sur_2},C^{sur}_{\pm(p,n)}),  \label{eq:paran2} 
\end{align}
where the $\omega$ functions in Eqs.~\eqref{eq:paranl} and~\eqref{eq:paran2} are defined as
\begin{align}
\omega _{n}(E,A^{sur},B^{sur},C^{sur})=A^{sur}\;\Theta \left(
X\right) \frac{X^{n}}{X^{n}+\left( B^{sur}\right) ^{n}}, \nonumber \\
\label{eq:omega}
\end{align}%
and $\Theta \left( X\right) $ is Heaviside's step function and $%
X=\left\vert E-\varepsilon_F\right\vert -C^{sur}$. 

The imaginary spin-orbit term in Eq.~\eqref{eq:imnl} has the same form as the real spin-orbit
term in Eq.~\eqref{eq:HFso},
\begin{align}
      W_{so}(\bm{r},\bm{r'};E)= \left( \frac{\hbar}{m_{\pi }c}\right) ^{2}
      W^{so}(E)\frac{1}{\tilde{r}}\frac{d}{d\tilde{r}}f(\tilde{r},r^{so}_{(p,n)},a^{so})\nonumber \\
      \times\bm{\ell}\cdot \bm{\sigma}
      H(\bm{s};\beta^{so}), 
   \label{eq:imag_so}
\end{align}
where the radial parameters for the imaginary component are the same as those used for the real part of the spin-orbit potential. 
It is important to note that $\textrm{Im}\Sigma_{so}$ grows with increasing $\ell$, and for large $\ell$ this can lead to an inversion of the sign of the self-energy, which results in negative occupation. While the
form of Eq.~\eqref{eq:HFso} suppresses this behavior, it is still not a proper solution. One must be careful that the magnitude of $W_{so}(E)$ does not exceed that of the volume and surface
components. As the imaginary spin-orbit component is
generally needed only at high energies, the form of Ref.~\cite{Mueller:2011} is employed, 
\begin{equation}
   W^{so}(E)= A^{so}  \frac{(E-\varepsilon_F)^4}{(E-\varepsilon_F)^4+(B^{so})^4} .
   \label{eq:ImSO}
\end{equation}%

With Eq.~\eqref{eq:ImSO}, all ingredients of the self energy have now been identified and their functional form described.
In addition to the Hartree-Fock contribution and the absorptive potentials we also include the
dispersive real part from all imaginary contributions according to Eq.~\eqref{eq:dispersion}.

\section*{Parameters}
\label{app:params}

\begin{table}[tb]
\caption{Parameter values for the isoscalar part of the potential. The table also contains the number of the
equation that defines each individual parameter.}

\label{Tbl:fixed}%
\begin{ruledtabular}

\begin{tabular}{ccSc} 
Parameter &  Unit & {\text{Value}} & Eq. No. \\

\hline
\multicolumn{4}{c}{Hartree-Fock} \\
\hline
$V^{HF}_{sym}$        &[MeV]      & 102.895   & (\ref{eq:HFv}) \\
$\beta^{vol_1}_{sym}$ &[fm]       &  1.552    &(\ref{eq:HFv}) \\
$\beta^{vol_2}_{sym}$ &[fm]       &  0.750    &(\ref{eq:HFv}) \\
$x_{sym}$ 	          &			  & 0.735	  & (\ref{eq:HFv}) \\
$\beta^{wb}$          & [fm]      &  0.937    & (\ref{eq:wb}) \\

\hline
\multicolumn{4}{c}{Spin-orbit} \\
\hline
$a^{so}$       &[fm]         &  0.700    & (\ref{eq:HFso}) \\
$\beta^{so}$   &[fm]         &  0.727    & (\ref{eq:HFso}) \\
$A^{so}$       &[MeV]        & -3.650    & (\ref{eq:ImSO}) \\
$B^{so}$       & [MeV]       &  208.000  &  (\ref{eq:ImSO})     \\

\hline
\multicolumn{4}{c}{Volume imaginary} \\
\hline
$a^{vol}_{+}$           &[fm] 		& 0.524 	& (\ref{eq:imnl}) \\ 
$a^{vol}_{-}$           &[fm] 		& 0.333 	& (\ref{eq:imnl}) \\ 
$B^{vol}_{+}$           &[MeV]		& 10.376	& (\ref{eq:volumeS}) \\
$B^{vol}_{-}$           &[MeV]		& 80.476	& (\ref{eq:volumeS}) \\
$\mathcal{E}^{vol}_{+}$ &[MeV]	    & 6.365	    & (\ref{eq:volumeS}) \\
$\mathcal{E}^{vol}_{-}$ &[MeV]		& 6.245	    & (\ref{eq:volumeS}) \\
$\mathbb{E}_{+}$        &[MeV]		& 15.831	& (\ref{eq:Wnmnl}) \\
$\mathbb{E}_{-}$        &[MeV]		& 99.122	& (\ref{eq:Wnmnl}) \\

\hline
\multicolumn{4}{c}{Surface imaginary} \\
\hline
$a^{sur}_{+sym}$     &[fm]    & 0.425 	 & (\ref{eq:imnl}) \\
$a^{sur}_{-sym}$     &[fm] 	  & 0.648 	 & (\ref{eq:imnl}) \\ 
$\beta^{sur0}_{+}$   &[fm]    & 1.404    & (\ref{eq:imnl}) \\ 
$\beta^{sur0}_{-}$   & [fm]   & 1.264    & (\ref{eq:imnl}) \\ 
$A^{sur0}_{+}$       & [MeV]  & 11.855   & (\ref{eq:paranl}) \\
$A^{sur0}_{-}$       & [MeV]  & 8.534    & (\ref{eq:paranl}) \\
$B^{sur0_1}_{+}$     & [MeV]  & 16.431   & (\ref{eq:paranl}) \\
$B^{sur0_1}_{-}$     & [MeV]  & 14.021   & (\ref{eq:paranl}) \\
$B^{sur0_2}_{+}$     & [MeV]  & 29.822   & (\ref{eq:paranl}) \\
$B^{sur0_2}_{-}$     & [MeV]  & 80.238   & (\ref{eq:paranl}) \\
$C^{sur0}_{+}$       & [MeV]  & 29.032   &  (\ref{eq:paranl}) \\
$C^{sur0}_{-}$       & [MeV]  & 9.500   &  (\ref{eq:paranl}) \\
\end{tabular}
\end{ruledtabular}
\end{table}

The parameters used for the symmetric part of the self-energy are presented in
Table~\ref{Tbl:fixed}. All asymmetric parameters are presented in
Table~\ref{Tbl:fitted}. 
There are 30 Lagrange-Legendre and Lagrange-Laguerre grid points used in the 
calculations~\cite{Baye_review,Baye:2010}. For $^{90}$Zr, the scaling parameter for the
Lagrange-Laguerre mesh points is $a_L=0.15$. The matching radius used for
scattering calculations is $a=12$~fm.

\begin{table}[tb]
\caption{Fitted parameter values for proton and neutron potentials in 
$^{90}$Zr. This table also lists the number of the equation that defines each individual parameter.}
\label{Tbl:fitted}%

\begin{ruledtabular}
\begin{tabular}{ccSSc} 
Parameter & Unit &  {\text{$p$ value}}    &  {\text{$n$ value}} & Eq. No. \\

\hline
\multicolumn{5}{c}{Hartree-Fock} \\
\hline

$r^{HFsym}_{(p,n)}$         & [fm]  &  1.146 & 1.130    & (\ref{eq:HFv}) \\
$a^{HFsym}_{(p,n)}$         & [fm]  & 0.660  & 0.700    & (\ref{eq:HFv}) \\
$V_{(p,n)}^{wb}$            & [MeV] & 5.580  &  2.107   & (\ref{eq:wb}) \\
$\rho_{(p,n)}^{wb}$         &[MeV]  & 0.548  &  0.800   & (\ref{eq:wb}) \\
$V^{HFasy}_{(p,n)}$         & [MeV] & 5.675  &  10.081  & (\ref{eq:HFv}) \\
$r^{HFasy}_{(p,n)}$         & [fm]  & 1.106  & 1.304    & (\ref{eq:HFv}) \\
$a^{HFasy}_{(p,n)}$         & [fm]  & 0.393  & 0.505    & (\ref{eq:HFv}) \\
$\beta^{vol_1asy}_{(p,n)}$  & [fm]  & 0.777  & 1.700    &(\ref{eq:HFv}) \\
$\beta^{vol_2asy}_{(p,n)}$  & [fm]  & 3.991  & 1.450    &(\ref{eq:HFv}) \\

\hline
\multicolumn{5}{c}{Spin-orbit} \\
\hline

$V^{so}_{(p,n)}$   &[MeV]     & 14.058  &  11.638  & (\ref{eq:HFso}) \\
$r^{so}_{(p,n)}$   &[fm]      & 1.165   &  1.158   & (\ref{eq:HFso}) \\ 

\hline
\multicolumn{5}{c}{Volume imaginary} \\
\hline
$\beta^{vol}_{+(p,n)}$ &[fm]  & 0.319   &  0.250 &(\ref{eq:imnl}) \\
$\beta^{vol}_{-(p,n)}$ &[fm]  & 1.143   &  1.143 &(\ref{eq:imnl}) \\
$r^{vol}_{+(p,n)}$     &[fm]  & 1.353   &  1.333 &(\ref{eq:imnl}) \\
$r^{vol}_{-(p,n)}$     &[fm]  & 1.129   & 1.198  & (\ref{eq:imnl}) \\
$A^{vol}_{+(p,n)}$     &[MeV] & 6.932	& 2.207	 & (\ref{eq:volumeS})\\
$A^{vol}_{-(p,n)}$     &[MeV] & 32.009	& 22.197 & (\ref{eq:volumeS})\\
$\alpha^{sym}_{(p,n)}$ &[fm]  & 0.084   & 0.358  & (\ref{eq:Wnmnl}) \\

\hline
\multicolumn{5}{c}{Surface imaginary} \\
\hline
$\beta^{sur}_{+(p,n)}$  &[fm]  &  0.209  & 2.150  & (\ref{eq:imnl}) \\
$\beta^{sur}_{-(p,n)}$  &[fm]  &  1.443  & 2.229  & (\ref{eq:imnl}) \\
$r^{sur0}_{+(p,n)}$     &[fm]  &  1.194  & 1.336  & (\ref{eq:imnl}) \\
$r^{sur0}_{-(p,n)}$     &[fm]  &  0.966  & 0.947  & (\ref{eq:imnl}) \\
$r^{sur}_{+(p,n)}$      &[fm]  &  1.199  & 0.850  & (\ref{eq:imnl}) \\
$r^{sur}_{-(p,n)}$      &[fm]  &  0.607  & 0.860  & (\ref{eq:imnl}) \\
$a^{sur}_{+(p,n)}$      &[fm]  &  0.536  & 0.370  & (\ref{eq:imnl}) \\
$a^{sur}_{-(p,n)}$      &[fm]  &  0.899  & 0.700  & (\ref{eq:imnl}) \\
$A^{sur}_{+(p,n)}$      &[MeV] &  6.000  & 5.488  & (\ref{eq:paran2}) \\
$A^{sur}_{-(p,n)}$      &[MeV] &  1.296  & 10.010 & (\ref{eq:paran2}) \\
$B^{sur_1}_{+(p,n)}$    &[MeV] &  28.701 & 25.139 & (\ref{eq:paran2}) \\
$B^{sur_1}_{-(p,n)}$    &[MeV] &  4.183  & 30.243 & (\ref{eq:paran2}) \\
$B^{sur_2}_{+(p,n)}$    &[MeV] &  24.502 & 24.398 & (\ref{eq:paran2}) \\
$B^{sur_2}_{-(p,n)}$    &[MeV] &  34.316 & 25.037 & (\ref{eq:paran2}) \\
$C^{sur}_{+(p,n)}$      &[MeV] &  20.734 & 12.082 & (\ref{eq:paran2}) \\
$C^{sur}_{-(p,n)}$      &[MeV] &  22.910 & 5.000  & (\ref{eq:paran2}) \\

\end{tabular}
\end{ruledtabular}
\end{table}

\newpage

\bibliographystyle{apsrev4-1}
\bibliography{Zr90}

\end{document}